\documentclass[journal]{IEEEtran}

\usepackage{amsmath,amsfonts,amsthm}
\usepackage{graphicx}
\usepackage{cite}
\usepackage{epstopdf}
\usepackage{float}
\usepackage{subfigure}
\usepackage{setspace}
\usepackage{stfloats}
\usepackage{url}

\begin{document}

\title{Jamming-assisted Eavesdropping over\\Parallel Fading Channels}

\author{Yitao~Han,
        Lingjie~Duan,~\IEEEmembership{Senior~Member,~IEEE,}
        and~Rui~Zhang,~\IEEEmembership{Fellow,~IEEE}
\thanks{Y. Han and L. Duan are with the Engineering Systems and Design Pillar, Singapore University of Technology and Design (e-mail: yitao\_han@mymail.sutd.edu.sg, lingjie\_duan@sutd.edu.sg).}
\thanks{R. Zhang is with the Department of Electrical and Computer Engineering, National University of Singapore (e-mail: elezhang@nus.edu.sg).}
}

\maketitle

\begin{abstract}

Unlike passive eavesdropping, proactive eavesdropping is recently proposed to use jamming to moderate a suspicious link's communication rate for facilitating simultaneous eavesdropping. This paper advances the proactive eavesdropping research by considering a practical half-duplex mode for the legitimate monitor (e.g., a government agency) and dealing with the challenging case that the suspicious link opportunistically communicates over parallel fading channels. To increase eavesdropping success probability, we propose cognitive jamming for the monitor to change the suspicious link's long-term belief on the parallel channels' distributions, and thereby induce it to transmit more likely over a smaller subset of unjammed channels with a lower transmission rate. As the half-duplex monitor cannot eavesdrop the channel that it is simultaneously jamming to, our jamming design should also control the probability of such ``own goal'' that occurs when the suspicious link chooses one of the jammed (uneavesdroppable) channels to transmit. We formulate the optimal jamming design problem as a mixed integer nonlinear programming (MINLP) and show that it is non-convex. Nevertheless, we prove that the monitor should optimally use the maximum jamming power if it decides to jam, for maximally reducing suspicious link's communication rate and driving the suspicious link out of the jammed channels. Then we manage to simplify the MINLP to integer programming and reveal a fundamental trade-off in deciding the number of jammed channels: jamming more channels helps reduce the suspicious link's communication rate for overhearing more clearly, but increases own goal probability and thus decreases eavesdropping success probability. Finally, we extend our study to the two-way suspicious communication scenario, and show there is another interesting trade-off in deciding the common jammed channels for balancing bidirectional eavesdropping performances. Numerical results show that our optimized jamming-assisted eavesdropping schemes greatly increase eavesdropping success probability as compared with the conventional passive eavesdropping.

\end{abstract}


\section{Introduction}

Security issues in wireless communication have drawn increasingly more attentions from both academia and industry. Due to the broadcast nature of wireless communication, its physical layer is vulnerable to eavesdropping (interception of confidential information) and jamming (interruption of legitimate transmission) \cite{zou2016sur}, and there are many works focusing on defence schemes against eavesdropping and jamming, such as secrecy beamforming \cite{li2011beamforming,amariucai2012half,mukherjee2011full}, channel-based secret key \cite{ren2011secret}, using cooperative networks to avoid eavesdropping \cite{wang2017anti} and using hopping to avoid jamming attacks \cite{navda2007using}. These works view eavesdropping or jamming as malicious attacks and assume all the communication links are rightful (see \cite{kashyap2004corr,liu2015disrupting,zhou2012pilot,kapetanovic2015physical,xiong2015energy,zou2015improving}). However, they overlook the emerging case that wireless links or devices can be established and used by criminals or terrorists to present severe public security threats.

With the fast development of wireless technologies and devices, user-controlled or infrastructure-free communications (e.g., ad hoc network and short-range communication) now become popular. For examples, mobile applications such as MeshMe and FireChat can network users in the vicinity with reliable mutual connection, and drones can take nice photos or videos and send back to their users. While providing great convenience to normal users, these new technologies and devices can be misused to commit crimes. Terrorists can use them to facilitate their plotting and acts, and spies can use them to send out commercial or military secrets. Since the data do not go through any public infrastructure under internet service providers (ISPs), it is difficult to be monitored by government surveillance program. As a result, prior methods (e.g., deploying dedicated wiretapping devices in network infrastructure) for eavesdropping infrastructure-based communications (e.g., cellular networks) no longer work. There is thus a growing need for authorized parties to develop new approaches to legitimately eavesdrop these infrastructure-free suspicious wireless communications. For example, in the USA, the National Security Agency has launched Terrorist Surveillance Program and aims to intercept all wireless devices \cite{TSP} to protect public security.

Traditionally, passive eavesdropping is used for such surveillance purpose but it does not provide good eavesdropping performance once the suspicious transmitter (ST) is far away from the monitor or hops to an undesirable channel. Recently, a novel approach called proactive eavesdropping via jamming is proposed in \cite{xu2017surveil}, \cite{xu2016proactive1}, where the legitimate monitor, ideally operating in full-duplex mode, uses jamming to moderate the suspicious communication rate for facilitating simultaneous eavesdropping. \cite{xu2017proactive1} extends this work by assuming the legitimate monitor's knowledge of full channel state information (CSI) and designs adaptive jamming power in each fading block. \cite{xu2017harq} studies the case that the suspicious link adopts HARQ-based communication. \cite{zhong2017multi} further considers that the monitor is equipped with multiple antennas to achieve more efficient jamming and better eavesdropping performance. \cite{zeng2016wireless} proposes another efficient eavesdropping method, where the monitor disguises as a fake relay to overhear the suspicious communication. These works largely assume that there is only one communication channel between the ST and suspicious receiver (SR), and they require the monitor to operate in interference-free full-duplex mode for enabling simultaneous jamming and eavesdropping. However in practice, full-duplex mode is difficult to implement, and self interference cancellation is hard to achieve as perfect \cite{riihonen2010residual} \cite{masmoudi2016self}. Rather, half-duplex mode is more widely used, and usually there is more than one channel for the suspicious link to communicate over.

\begin{figure}[!t]
\centering
\includegraphics[width=3.0in]{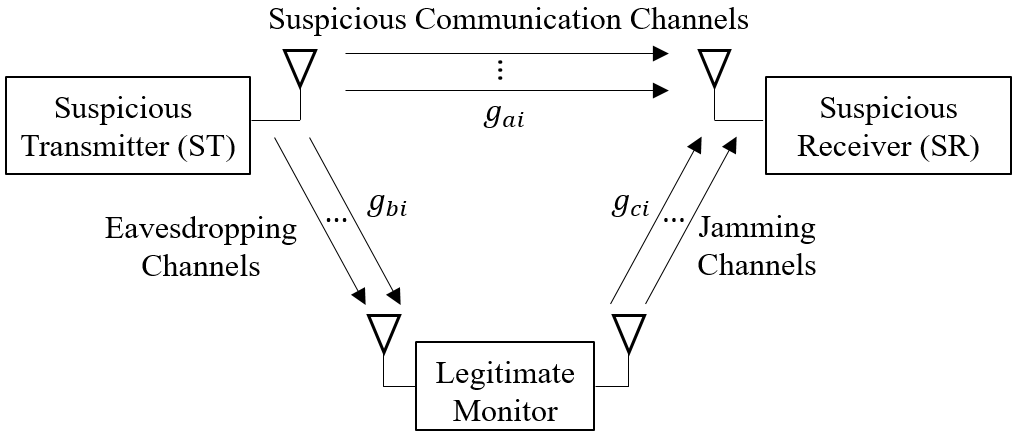}
\caption{System model of jamming-assisted eavesdropping over the suspicious link (suspicious transmitter to receiver).}
\label{fig1}
\end{figure}

In this paper, we study a practical wireless surveillance scenario: a half-duplex legitimate monitor eavesdrops from a suspicious communication link over parallel independent Rayleigh fading channels. As shown in Fig.~\ref{fig1}, at the beginning of each fading block, based on the conditions of the parallel independently fading channels, the ST hops to the best one for transmission in this fading block.\footnote{This approach of channel hopping and selection is commonly used in the multi-channel scenario to enhance wireless security against eavesdropping and jamming (e.g., \cite{navda2007using}). In the future work, we will further consider the case that the ST uses multiple channels to transmit, e.g., by deploying waterfilling-based power allocation over multiple channels, which will bring more challenges to legitimate eavesdropping.} Here we assume a typical delay-sensitive application (e.g., video talk) on the suspicious link, i.e., the transmitter adjusts its transmission rate by maintaining a certain target outage probability at the receiver \cite{xu2016proactive1}. Usually, the monitor is far away from the ST to stay undetected, which makes the traditional passive eavesdropping difficult or even infeasible. Under this challenging setup, the monitor can deliberately send jamming signals to the SR to induce the ST to transmit more likely over a smaller subset of unjammed channels with a lower transmission rate, so that the monitor can still eavesdrop effectively.

To avoid getting exposed, the monitor will not change its jamming power and jammed channels over time, it just disguises itself as a normal device-to-device (D2D) user in sharing the network, by sending randomly modulated messages over fixed channels with fixed power. The ST/SR is aware of its co-existence in the same network by updating the long-term belief of the parallel channels' distributions, but does not consider it as a jammer. If the monitor keeps changing its jammed channels or jamming power, then it is no longer like a normal user and will cause the ST/SR's suspicion to hop over channels to increase resilience in a game theoretic setting or directly stop transmitting any message as in \cite{navda2007using} \cite{li2010dogfight1} \cite{li2011dogfight2}.

The key novelty and main contributions of this paper are summarized as follows.

\begin{itemize}
  \item \emph{Novel jamming-assisted eavesdropping approach over parallel fading channels:}\ To our best knowledge, this is the first paper studying wireless surveillance of parallel fading channels via a half-duplex monitor. The monitor uses jamming to change the suspicious link's long-term belief on the parallel channels' distributions, and thereby induce it to transmit more likely in a smaller subset of unjammed channels with a lower transmission rate for higher eavesdropping success probability. For practical concern, we consider a challenging case that the monitor has no instantaneous CSI of any suspicious link channels, and the monitor in half-duplex mode cannot eavesdrop a channel that it is simultaneously jamming to.
  \item \emph{Joint optimization of jamming power and number of jammed channels:} We formulate the problem for optimal jamming design over parallel fading channels as a mixed integer nonlinear programming (MINLP) and show it is non-convex. Nevertheless, we prove that the monitor should use the maximum jamming power if it decides to jam. Then we manage to simplify the MINLP to integer programming and further show that there is a fundamental trade-off in deciding the number of jammed channels: jamming more channels helps reduce the suspicious communication rate for overhearing more clearly, but at the risk that the ST is more likely to choose among the jammed channels to transmit and as a result cannot be overheard.
  \item \emph{Jamming-assisted eavesdropping over two-way communications:} We extend the model to consider the two-way communications of the suspicious link. As the monitor cannot change its jamming strategy to avoid getting exposed, it needs to jam the same subset of channels for both communication directions. To decide the optimal number of jammed channels, we show there is another trade-off to balance the eavesdropping performances of the two-way communications.
  \item \emph{Performance evaluation:} Numerical results show that our jamming-assisted eavesdropping schemes achieve great performance gain over conventional passive eavesdropping. We also show that the monitor will perform passive eavesdropping only when it is close to the ST, and will jam increasingly more channels when it is moving away from the ST to SR, due to deteriorating eavesdropping channels and improving jamming channels.
\end{itemize}

The rest of this paper is organized as follows. In Section II, we present the system model and formulate the legitimate monitor's proactive eavesdropping design problem. In Section III, we solve the eavesdropping optimization problem in the special two-channel case to gain useful insights. In Section IV, we extend to multi-channel case and show an interesting trade-off in deciding the number of jammed channels, for eavesdropping success probability maximization. In Section V, we further consider jamming-assisted eavesdropping over two-way communications. In Section VI, we provide more numerical results to evaluate our jamming-assisted eavesdropping approach. Finally, we conclude this paper in Section VII.

\section{System model and problem formulation}

\begin{table*}[ht]
\newcommand{\tabincell}[2]{\begin{tabular}{@{}#1@{}}#2\end{tabular}}
\caption{Symbols and their physical meanings}
\centering
\begin{tabular}{|c|c|}
\hline
Symbols & Physical meanings\\
\hline
\hline
$N$ & Total number of parallel channels.\\
\hline
$n$ & Number of jammed channels.\\
\hline
$g_{ai}$, $g_{bi}$, $g_{ci}$ & \tabincell{c}{Channel power gain of the suspicious link with mean $1/\lambda_a$, eavesdropping link with mean $1/\lambda_b$,\\ jamming link with mean $1/\lambda_c$, on channel $i$.}\\
\hline
$P$ & ST's transmitting power.\\
\hline
$Q_i$ & Monitor's jamming power on channel $i$.\\
\hline
$Q_{\max}$ & Jamming power budget of the monitor.\\
\hline
$\sigma_a^2$, $\sigma_b^2$, $\sigma_c^2$ & \tabincell{c}{Noise power level at the SR, the monitor, the ST.}\\
\hline
$R^{I}$, $R^{II}$ & \tabincell{c}{Suspicious transmission rate of passive eavesdropping, jamming-assisted eavesdropping.}\\
\hline
$\delta$ & \tabincell{c}{Target outage probability at the SR.}\\
\hline
$\rho$ & \tabincell{c}{Own goal probability (probability that one of the jammed channels is chosen by the ST for transmission).}\\
\hline
$\varphi ^{I}$, $\varphi ^{II}$ & \tabincell{c}{Eavesdropping success probability of passive eavesdropping, jamming-assisted eavesdropping.}\\
\hline
\end{tabular}
\vspace{-0.5em}
\end{table*}

As shown in Fig.~\ref{fig1}, the ST communicates with the SR over $N\geq 2$ parallel channels with independent Rayleigh fading, and there is a legitimate monitor aiming to eavesdrop their communications. We consider a quasi-stationary system model, where the monitor has sufficient time (before the ST and SR's movement to another locations) to learn the global channel distribution information (CDI) and launch jamming to eavesdrop from the suspicious link's transmission. The ST and SR are both equipped with one antenna, while the legitimate monitor is purposely equipped with two antennas, one for receiving (eavesdropping) and the other for transmitting (jamming). In order to characterize the fundamental performance limit of this jamming-assisted eavesdropping approach, we assume the encryption method used by the ST/SR is known to the monitor beforehand (e.g. via eavesdropping the related encryption codebook). Note that the focus of this work is on decoding the message instead of decrypting the message itself. The monitor disguises itself as a normal user in sharing the same set of channels with the suspicious link, and operates at half-duplex mode, which means it cannot eavesdrop the channel that it is jamming to. Thus, the monitor will not jam all $N$ channels, otherwise, it overhears nothing. There are two eavesdropping schemes to investigate and compare:

\begin{itemize}
  \item \emph{Scheme I (passive eavesdropping):}\ The legitimate monitor performs passive eavesdropping over all $N$ channels while jamming no channels. This is also a benchmark case for our proposed jamming-assisted eavesdropping to compare with.
  \item \emph{Scheme II (jamming-assisted eavesdropping):}\ The legitimate monitor performs jamming-assisted eavesdropping by jamming $n$ channels and eavesdropping from the rest $N-n$ channels, where $1\leq n\leq N-1$.
\end{itemize}

We consider a block fading model, where the channel stays unchanged in each fading block and may vary over different fading blocks. As shown in Fig.~\ref{fig1}, we respectively denote the channel power gains of the suspicious communication link, eavesdropping link (from the ST to the monitor) and jamming link (from the monitor to the SR) on channel $i\in\{1,\cdots,N\}$ as $g_{ai}$, $g_{bi}$ and $g_{ci}$. By considering independent Rayleigh fading, $g_{ai}$, $g_{bi}$ and $g_{ci}$ are modelled as independent exponentially distributed random variables with mean $1/\lambda_a$, $1/\lambda_b$ and $1/\lambda_c$, respectively, with $i\in\{1,\cdots,N\}$. This suggests that all the suspicious link channels ($g_{ai}$, $\forall i$) are independent and identically distributed (i.i.d.), so are the eavesdropping channels ($g_{bi}$, $\forall i$) and the jamming channels ($g_{ci}$, $\forall i$). Hence in Scheme II, in statistical sense it does not matter which channels to jam given the jammed channel number by the monitor (who does not know the instantaneous CSI of any suspicions link channels). Thus, without loss of generality, we assume the half-duplex monitor picks the first $n$ out of $N$ channels to jam, and eavesdrops from the rest $N-n$ channels. We assume that the monitor only knows the global CDI ($1/\lambda_a$, $1/\lambda_b$ and $1/\lambda_c$), which can be obtained by the monitor through long-term observation as mentioned earlier.

On the other hand, we consider that the ST knows the CSI of all the suspicious communication channels (i.e., $g_{ai}$'s instantaneous values). The ST transmits at a fixed power $P$ and keeps hopping to the best channel for transmission in each fading block. For ease of reading, Table I summarizes the main symbol notations used in this paper and their physical meanings.

\subsection{Monitor's expected performance of the suspicious link without or with jamming}

Both the signal sent by the ST and the jamming signal sent by the monitor are assumed to be circularly symmetric complex Gaussian (CSCG) random variables. This is because that CSCG message will achieve channel capacity given CSCG noise, and monitor's CSCG jamming signal will achieve the best jamming effect \cite{kashyap2004corr}.

In Scheme I, the monitor does not jam, and the achievable rate of suspicious communication on channel $i\in\{1,\cdots,N\}$ is $\log_2(1+\frac{g_{ai}P}{\sigma_a^2})$ in bits/second/Hertz (bps/Hz), where $\sigma_a^2$ denotes the noise power at the SR. The monitor expects the signal-to-noise ratio (SNR) $\frac{g_{ai}P}{\sigma_a^2}$ at the SR on the $i^{th}$ channel is a random variable, with cumulative distribution function (CDF) given by
\begin{equation}
\begin{aligned}
\mathbb{P}\bigg(\frac{g_{ai}P}{\sigma_a^2}\leq \gamma\bigg)=1-e^{-\frac{\lambda_{a}\sigma_a^2}{P}\gamma}, \ \ \gamma\geq 0.
\label{ii1}
\end{aligned}
\end{equation}

In Scheme II, the monitor jams by allocating $Q_i\geq 0$ power to channel $i\in\{1,\cdots,N\}$, and thus the achievable rate of suspicious communication channel $i$ is $\log_2(1+\frac{g_{ai}P}{g_{ci}Q_i+\sigma_a^2})$. The monitor expects the signal-to-interference-plus-noise ratio (SINR) at the SR on the $i^{th}$ channel is a random variable, with CDF given by the following lemma.

\underline{\emph{Lemma}} \ \emph{2.1:} \ In Scheme II, the CDF of SINR $\frac{g_{ai}P}{g_{ci}Q_i+\sigma_a^2}$ at the SR under jamming is given by
\begin{equation}
\begin{aligned}
\mathbb{P}\bigg(\frac{g_{ai}P}{g_{ci}Q_i+\sigma_a^2}\leq \gamma\bigg) = 1-\frac{\lambda_cPe^{-\frac{\lambda_a\sigma_a^2}{P}\gamma}}{\lambda_cP+\lambda_a Q_i \gamma}, \ \ \ \gamma\geq 0.
\label{ii2}
\end{aligned}
\end{equation}

\begin{IEEEproof}
See Appendix A.
\end{IEEEproof}

Based on \eqref{ii1} and \eqref{ii2} under the two eavesdropping schemes, we are ready to formulate the monitor's design objective.

In Scheme I, the legitimate monitor performs passive eavesdropping. The ST will choose the best channel, i.e., the one with the highest SNR, in each fading block. From \eqref{ii1}, the monitor's expected CDF of the maximum SNR $\gamma^{I}$ at the SR among all the channels is
\begin{equation}
\begin{aligned}
\mathbb{P}\bigg(\gamma^{I}&=\max\bigg\{\frac{g_{a1}P}{\sigma_a^2},\cdots,\frac{g_{aN}P}{\sigma_a^2}\bigg\}\leq \gamma\bigg) \\
         &=\Big(1-e^{-\frac{\lambda_a\sigma_a^2}{P}\gamma}\Big)^{N}, \ \ \ \ \gamma\geq 0.
\label{ii3}
\end{aligned}
\end{equation}

We consider a typical delay-sensitive transmission model for the suspicious link, where the ST adjusts its transmission rate to keep a target outage probability $\delta$ at the SR. Only when the transmission rate $R^{I}$ is no larger than the achievable rate of the best suspicious communication channel $r_{a}^{I} = \log_2(1+\gamma^{I})$, the SR can successfully decode the delay-sensitive message. Thus we have
\begin{equation}
\begin{aligned}
\mathbb{P}(r_{a}^{I}<R^{I})=\delta,
\label{ii4}
\end{aligned}
\end{equation}which yields the suspicious transmission rate $R^{I}$ as
\begin{equation}
\begin{aligned}
R^{I} = \log_2\bigg(1+\frac{P}{\lambda_a\sigma_a^2}\ln\frac{1}{1-\delta^{\frac{1}{N}}}\bigg).
\label{ii5}
\end{aligned}
\end{equation}

In Scheme II, the legitimate monitor performs jamming-assisted eavesdropping and jams $n\geq 1$ channels. Under jamming, the ST will choose the channel with the highest SINR among the $n$ jammed channels or the channel with the highest SNR among the remaining $N-n$ channels without jamming in each fading block. Note that ST's chosen channel may still be a jammed channel by the monitor due to independent channel fading. From \eqref{ii1} and \eqref{ii2}, the monitor's expected CDF of the maximum SINR or SNR $\gamma^{II}$ at the SR among all the channels is
\begin{equation}
\begin{aligned}
&\mathbb{P}\bigg(\gamma^{II}=\max\bigg\{\frac{g_{a1}P}{g_{c1}Q_1+\sigma_a^2},\cdots,\frac{g_{an}P}{g_{cn}Q_n+\sigma_a^2},\\ &\ \ \ \ \ \ \ \ \ \ \ \ \ \ \ \ \ \ \ \ \ \ \ \ \ \ \ \ \ \ \ \ \ \frac{g_{a(n+1)}P}{\sigma_a^2}, \cdots,\frac{g_{aN}P}{\sigma_a^2}\bigg\}\leq \gamma\bigg)\\
&=\Big(1-e^{-\frac{\lambda_a\sigma_a^2}{P}\gamma}\Big)^{N-n}\prod_{i=1}^n\bigg(1-\frac{\lambda_cPe^{-\frac{\lambda_a\sigma_a^2}{P}\gamma}}{\lambda_cP+\lambda_a Q_i \gamma}\bigg), \ \ \gamma\geq 0.
\label{ii6}
\end{aligned}
\end{equation}

To maintain target outage probability $\delta$ at the SR, the ST sets the transmission rate $R^{II}$ to ensure
\begin{equation}
\begin{aligned}
\mathbb{P}(\log_2(1+\gamma^{II})<R^{II})=\delta,
\label{ii7}
\end{aligned}
\end{equation}which yields
\begin{equation}
\begin{aligned}
\Big(\!1\!-e^{-\frac{\lambda_a\sigma_a^2}{P}(2^{R^{II}}\!\!-1)}\!\Big)^{N-n} \!\prod\limits_{i=1}^n\bigg(\!1\!-\!\frac{\lambda_cPe^{-\frac{\lambda_a\sigma_a^2}{P}(2^{R^{II}}\!\!-1)}}{\lambda_cP\!+\!\lambda_a Q_i (2^{R^{II}}\!\!-\!\!1)}\!\bigg)\!=\!\delta.
\label{ii8}
\end{aligned}
\end{equation}

With its jamming power allocations $\{Q_i\}_{i=1}^n$, the monitor believes the ST will use rate $R^{II}$ to transmit. Note that there is no closed-form solution $R^{II}$ to equation \eqref{ii8}.

\subsection{Monitor's problem formulation for jamming-assisted eavesdropping}

The legitimate monitor aims to maximize the eavesdropping success probability on the suspicious communication, which is the percentage of fading blocks it can successfully decode.

In Scheme I, assuming the suspicious communication is on the $i^{th}$ channel in a certain fading block, only when the achievable rate of the $i^{th}$ eavesdropping channel is no smaller than the suspicious transmission rate $R^I$ in \eqref{ii5}, the monitor can successfully eavesdrop in this fading block. Thus, the eavesdropping success probability $\varphi ^{I}$ under Scheme I (passive eavesdropping) is
\begin{equation}
\begin{aligned}
\varphi ^{I} \!=\! 1\!-\!\mathbb{P}\bigg\{r_{b}\!=\!\log_2\Big(1\!+\!\frac{g_{bi}P}{\sigma_b^2}\Big)\!<\!R^{I}\bigg\}\!=\! e^{-\frac{\lambda_b\sigma_b^2}{P}(2^{R^{I}}-1)},
\label{ii9}
\end{aligned}
\end{equation}where $\sigma_b^2$ denotes the noise power at the legitimate monitor.

In Scheme II, since all the suspicious communication channels have independent fading, it is possible that a jammed channel is still chosen by the ST, and in this case the half-duplex monitor cannot eavesdrop anything. We define the probability that any jammed channel is chosen by the ST (i.e., own goal probability of the monitor's self-jamming) as $\rho\in[0,1]$ and we will detail its analysis later in Sections III and IV.

Now, assuming the suspicious communication is on the $i^{th}$ channel in a certain fading block, only when none of the jammed channels is chosen by the ST, and the achievable rate of the $i^{th}$ eavesdropping channel is no smaller than the suspicious transmission rate $R^{II}$ in \eqref{ii8}, the monitor can successfully eavesdrop. Thus, the eavesdropping success probability $\varphi ^{II}$ under Scheme II (jamming-assisted eavesdropping) is
\begin{equation}
\begin{aligned}
&\varphi ^{II}\!(n,\!Q_1,\!\cdots\!,\!Q_n\!)\! =\! (1\!-\!\rho(n,\!Q_1,\!\cdots\!,\!Q_n)) \\ & \ \ \ \ \ \ \ \times\!\bigg(\!1\!-\!\mathbb{P}\bigg\{\!r_{b}\!=\!\log_2\Big(\!1\!+\!\frac{g_{bi}P}{\sigma_b^2}\Big)\!<\!R^{II}(n,\!Q_1,\!\cdots\!,\!Q_n)\!\bigg\}\bigg) \\ & \ \ \ \ \ \ \ =(1-\rho(n,\!Q_1,\!\cdots\!,\!Q_n))e^{-\frac{\lambda_b\sigma_b^2}{P}(2^{R^{II}(n\!,Q_1\!,\cdots\!,Q_n\!)}-1)}.
\label{ii10}
\end{aligned}
\end{equation}

From \eqref{ii9} and \eqref{ii10}, we can formulate the optimization problem for jamming design as a mixed integer nonlinear programming (MINLP), given by
\begin{equation}
\begin{aligned}
(\mathrm{P1}): &\max_{n,Q_1,\cdots,Q_n} \ \max \{\varphi ^{I},\varphi ^{II}(n,Q_1,\cdots,Q_n)\}\\
& \ \mathrm{s.t.} \ 0< \sum_{i=1}^{n} Q_i\leq Q_{\max},\\
& \ \ \ \ \ \ 1\leq n\leq N-1, n\in\mathbb{Z}.\nonumber
\label{p1}
\end{aligned}
\end{equation}

Note that $\varphi^I$ does not depend on $n$ or $Q_i$ and it is a constant. In other words, the monitor only jams if the optimized $\varphi ^{II}(n,Q_1,\cdots,Q_n)$ is larger than $\varphi^I$.

The joint optimization of $n$ and $Q_i$'s in problem ($\mathrm{P1}$) is difficult due to discrete $n$ and non-concave objective $\varphi ^{II}(n,Q_1,\cdots,Q_n)$. In the next section, we will first look into the two-channel case ($N=2$) to simplify this problem and provide tractable analysis and clean insights. We will generalize the results to the multi-channel case in Section IV.

\section{Optimal jamming-assisted eavesdropping over two channels}

If there are only two parallel channels ($N=2$), in jamming-assisted eavesdropping approach, the half-duplex legitimate monitor will only jam one channel (i.e., $n=1$), otherwise, both channels are jammed and it cannot overhear anything due to own goal. Its eavesdropping success probability $\varphi^{II}$ under our jamming-assisted eavesdropping in \eqref{ii10} now only depends on jamming power $Q$, i.e.,
\begin{equation}
\begin{aligned}
\varphi^{II}(Q)  = (1-\rho(Q))e^{-\frac{\lambda_b\sigma_b^2}{P}(2^{R^{II}(Q)}-1)},
\label{iii11}
\end{aligned}
\end{equation}and (P1) is simplified to
\begin{equation}
\begin{aligned}
(\mathrm{P2}): &\max_{Q} \ \max \{\varphi ^{I},\varphi ^{II}(Q)\}\\
& \ \mathrm{s.t.} \ 0< Q\leq Q_{\max}. \nonumber
\label{p2}
\end{aligned}
\end{equation}

\underline{\emph{Proposition 3.1:}} \ In the two-channel case, the own goal probability $\rho(Q)$ in \eqref{iii11} due to self-jamming at the monitor is given by
\begin{equation}
\begin{aligned}
\rho(Q) = -\frac{\lambda_c \sigma_a^2}{Q}e^{\frac{2\lambda_c \sigma_a^2}{Q}}\text{Ei}(-\frac{2\lambda_c\sigma_a^2}{Q}),
\label{iii12}
\end{aligned}
\end{equation}where $\text{Ei}(\cdot)$ is the exponential integral function [25, Eq. $8.21$]. As jamming power $Q$ increases, $\rho(Q)$ decreases. This is because a higher jamming power helps drive the suspicious link to the other unjammed (eavesdropped) channel. More specifically, we have $\lim_{Q\rightarrow 0^+} \rho(Q)=1/2$ due to trivial jamming effect on changing the ST's belief of channel distributions, and the suspicious link is equally likely to choose both i.i.d. channels to transmit. Moreover, $\lim_{Q\rightarrow \infty} \rho(Q)=0$, since the jammed channel will never be chosen by the suspicious link under infinite jamming power.

\begin{IEEEproof}
See Appendix C, where we choose $N=2$ and $n=1$. By substituting $Q=0^+$ and $Q=\infty$ into \eqref{iii12}, we can derive the two limits for $\rho(Q)$.
\end{IEEEproof}

\begin{IEEEproof}
\begin{figure}[!t]
\centering
\includegraphics[width=3.0in]{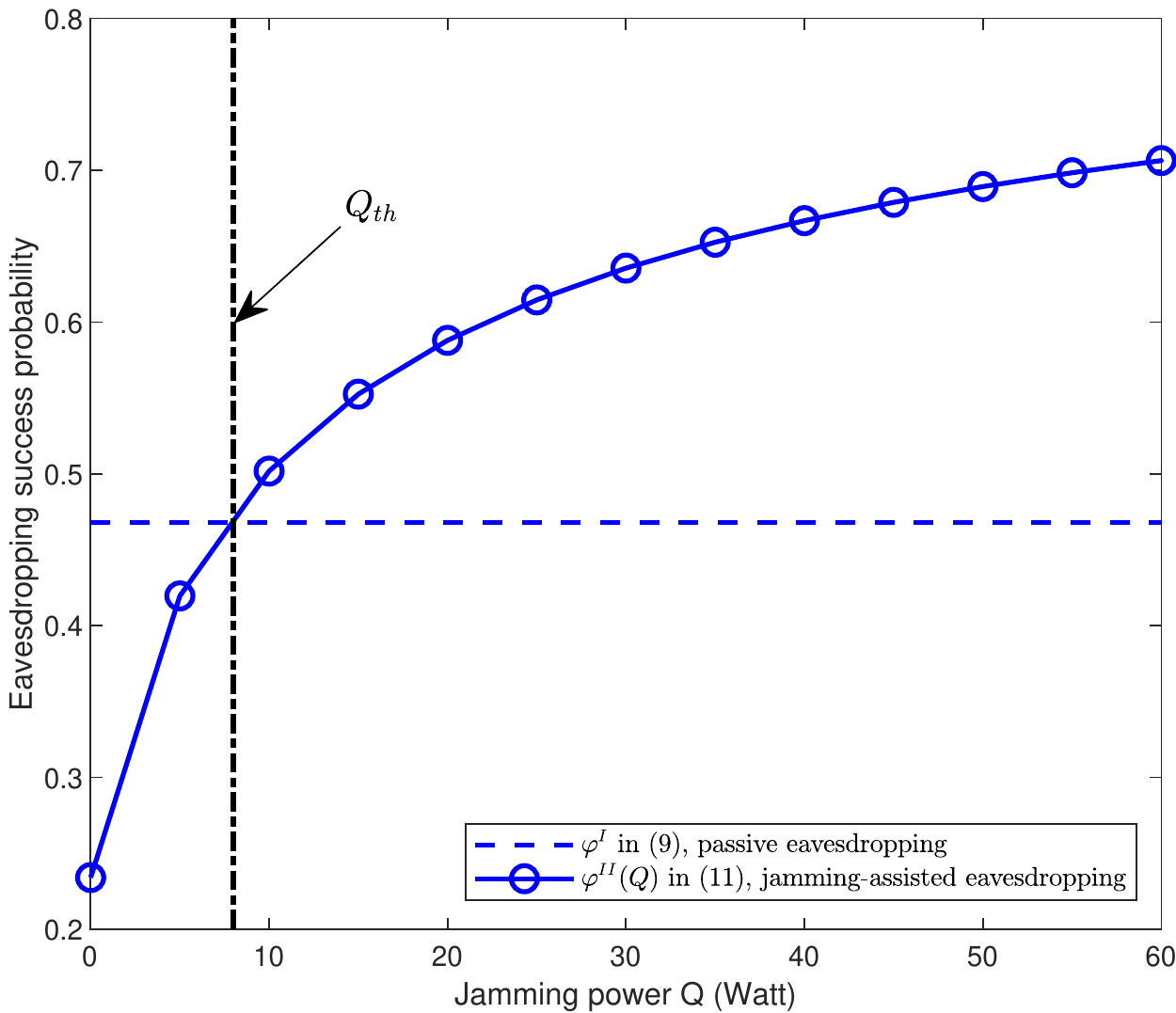}
\caption{Eavesdropping success probability versus jamming power at the monitor, where $N=2$, $\lambda_a=1$, $\lambda_b=\lambda_c=3$, $P=10$ dB, $\sigma_a^2=\sigma_b^2=1$ and $\delta=0.05$.}
\label{fig2}
\vspace{-0.8em}
\end{figure}

Next, we analyze the non-outage probability at the monitor $e^{-\lambda_b\sigma_b^2(2^{R^{II}(Q)}-1)/P}$, a part of \eqref{iii11}, which is a function of $R^{II}(Q)$. By simplifying \eqref{ii8} under $N=2$ and $n=1$, we have the following results.

\underline{\emph{Proposition 3.2:}} \ In the two-channel case, the suspicious link's transmission rate $R^{II}(Q)$ in \eqref{iii11} is the unique solution to

\begin{equation}
\begin{aligned}
\Big(\!1\!-\!e^{-\frac{\lambda_a\sigma_a^2}{P}(2^{R^{II}(Q)}-1)}\!\Big) \!\bigg(\!1\!-\!\frac{\lambda_cPe^{-\frac{\lambda_a\sigma_a^2}{P}(2^{R^{II}(Q)}-1)}}{\lambda_cP\!+\!\lambda_a Q (2^{R^{II}(Q)}\!-\!1)}\!\bigg)\!=\!\delta.
\label{iii13}
\end{aligned}
\end{equation}

As jamming power $Q$ increases, $R^{II}(Q)$ decreases. This is because the ST faces more noisy channel, and has to transmit at a lower rate in order to maintain target outage probability $\delta$. More specifically, $\lim_{Q\rightarrow 0^+}R^{II}(Q)=\log_2\big(1+P\ln(1-\delta^{\frac{1}{2}})^{-1}/\lambda_a\sigma_a^2\big)$,
which equals to $R^I$ in \eqref{ii5} with $N=2$ under passive eavesdropping, due to trivial jamming effect on the suspicious communication. Moreover, $\lim_{Q\rightarrow\infty}R^{II}(Q)=\log_2\big(1+P\ln(1-\delta)^{-1}/\lambda_a\sigma_a^2\big)$,
which equals to $R^I$ in \eqref{ii5} with $N=1$ under the passive eavesdropping, since the jammed channel will never be chosen.

\begin{IEEEproof}
By substituting $N=2$ and $n=1$ into \eqref{ii8}, we have \eqref{iii13}. Denote the left-hand-side (LHS) of \eqref{iii13} to be $g(R^{II}(Q),Q)$. According to the implicit function theorem, we have
\begin{equation}
\begin{aligned}
\frac{dR^{II}(Q)}{dQ} = -\frac{\partial g(R^{II}(Q),Q)/\partial Q}{\partial g(R^{II}(Q),Q)/\partial R^{II}(Q)}<0,\nonumber
\end{aligned}
\end{equation}
thus $R^{II}(Q)$ monotonically decreases as $Q$ increases. By substituting $Q=0^+$ and $Q=\infty$ into \eqref{iii13}, we can derive the two limits for $R^{II}(Q)$.
\end{IEEEproof}

\underline{\emph{Theorem 3.1:}} \ We denote $Q_\text{th}$ as the unique solution to
\begin{equation}
\begin{aligned}
\Big(1-\rho(Q_\text{th})\Big)e^{-\frac{\lambda_b\sigma_b^2}{P}(2^{R^{II}(Q_\text{th})}-1)}=e^{-\frac{\lambda_b\sigma_b^2}{P}(2^{R^{I}}-1)}.
\label{iii14}
\end{aligned}
\end{equation}If $Q_{\max}>Q_\text{th}$, the legitimate monitor will jam with the maximum power $Q_{\max}$, otherwise it will perform passive eavesdropping without jamming (as illustrated in Fig.~\ref{fig2}).

According to Propositions 3.1 and 3.2, $\varphi ^{II}(Q)$ in \eqref{iii11} monotonically increases with $Q$, thus the monitor will use the maximum jamming power $Q_{\max}$ if it decides to jam, while $\varphi ^{I}$ in \eqref{ii9} is a constant regardless of $Q$.

When jamming power $Q$ is close to zero,
\begin{equation}
\begin{aligned}
\lim_{Q\rightarrow 0^+}\varphi ^{II}(Q)=\frac{1}{2}e^{-\frac{\lambda_b\sigma_b^2}{P}(2^{R^{I}}-1)}< \varphi ^{I}.\nonumber
\end{aligned}
\end{equation}

On the other hand, when jamming power goes to infinity, $\lim_{Q\rightarrow\infty}R^{II}(Q)=R^{I}|_{(N=1)}<R^{I}$, hence
\begin{equation}
\begin{aligned}
\lim_{Q\rightarrow \infty}\varphi ^{II}(Q)=e^{-\frac{\lambda_b\sigma_b^2}{P}(2^{R^{I}|_{(N=1)}}-1)}>\varphi ^{I}.\nonumber
\end{aligned}
\end{equation}

Given that when jamming power goes to zero, the performance of jamming-assisted eavesdropping is worse than passive eavesdropping, while the performance of jamming-assisted eavesdropping increases with jamming power, and eventually when jamming power goes to infinity, becomes better than passive eavesdropping, it follows that there exists a unique intersection point between $\varphi ^{I}$ and $\varphi ^{II}(Q)$ at point $Q=Q_{\text{th}}$, which is given in \eqref{iii14}. If $Q_{\max}>Q_\text{th}$, the monitor will jam with the maximum jamming power $Q_{\max}$, otherwise it will perform passive eavesdropping to obtain a greater $\varphi^I$ than $\varphi^{II}(Q_{\max})$.
\end{IEEEproof}

As we can see from above, in jamming-assisted eavesdropping, jamming with a higher power helps reduce own goal probability $\rho(Q)$ and transmission rate $R^{II}(Q)$ at the same time. Thus, by using up jamming power budget $Q_{\max}$, the monitor can achieve the maximum eavesdropping success probability.

\section{Optimal jamming-assisted eavesdropping over multiple channels}

In this section, we consider the general case with multiple i.i.d. fading channels for the optimal jamming design, and need to further decide how many channels to jam. Similar to the $N=2$ case in Theorem 3.1, we also expect to jam with the maximum power $Q_{\max}$ in the general case of $N\geq 2$ channels if the monitor decides to jam. More specifically, we have the following result.

\underline{\emph{Proposition 4.1:}} \ Given that $n\geq1$ of $N$ channels are jammed, the monitor should allocate all the jamming power over $n$ jammed channels equally, i.e., $Q_i^*=Q_{\max}/n$, $\forall i\in\{1, \cdots, n\}$.

\begin{IEEEproof}
See Appendix B.
\end{IEEEproof}

Thanks to Proposition 4.1, we know that the monitor will evenly allocate all the jamming power over the jammed channels, as a result the eavesdropping success probability $\varphi^{II}$ under jamming-assisted eavesdropping in \eqref{ii10} only depends on the number of jammed channels $n$, i.e.,
\begin{equation}
\begin{aligned}
\varphi^{II}(n)  = (1-\rho(n))e^{-\frac{\lambda_b\sigma_b^2}{P}(2^{R^{II}(n)}-1)}.
\label{iv15}
\end{aligned}
\end{equation}

Thus, we manage to simplify the non-convex MINLP in (P1) to the following single-variable problem:
\begin{equation}
\begin{aligned}
(\mathrm{P3}): &\max_{n} \ \max \{\varphi ^{I},\varphi ^{II}(n)\}\\
&\ \mathrm{s.t.} \ 1\leq n\leq N-1, n\in\mathbb{Z}.\nonumber
\label{p3}
\end{aligned}
\end{equation}

As $n$ is an integer, problem ($\mathrm{P3}$) is an integer programming problem and still difficult to solve analytically. Next, we will analyze the monotonic properties of own goal probability $\rho(n)$ and suspicious link's transmission rate $R^{II}(n)$ with respect to $n$ in the objective, to understand the key insights and solve (P3).

\underline{\emph{Proposition 4.2:}} \ In the general multi-channel case, the own goal probability $\rho(n)$ at the monitor in \eqref{iv15} is given by
\begin{equation}
\begin{aligned}
\rho(n)\! =\! \frac{(N\!-\!n)\lambda_a\sigma_a^2}{P}\sum_{i=0}^{N\!-n\!-\!1}\!\sum_{j=1}^{n}\!{{N\!-\!n\!-\!1} \choose i}\!{n \choose j}\!(-1)^{i+j+1}\\ \times\bigg(\!\frac{\lambda_aQ_{\max}}{n\lambda_cP}\!\bigg)^{-j}\!\!\bigg(\!(1\!+\!i\!+\!j)\frac{\lambda_a\sigma_a^2}{P}\!\bigg)^{j-1}e^{\frac{(1+i+j)n\lambda_c\sigma_a^2}{Q_{\max}}}
\!\!\!\\
\times \Gamma\bigg(\!1\!-\!j,\frac{(1\!+\!i\!+\!j)n\lambda_c\sigma_a^2}{Q_{\max}}\!\bigg),
\label{iv16}
\end{aligned}
\end{equation}where $\Gamma(\cdot,\cdot)$ is the upper incomplete gamma function [25, Eq. $8.35$]. Further, $\rho(n)$ increases as $n$ increases.

\begin{IEEEproof}
See Appendix C.
\end{IEEEproof}

Here, $\rho(n)$ increases with $n$ due to two reasons. First, more channels are jammed and potentially they can be selected by the suspicious link for transmission. Second, the jamming power on each jammed channel weakens as $n$ increases given the total jamming power budget, and thus each jammed channel is more likely to be chosen by the suspicious link. Thus, the suspicious link is more likely to transmit on the jammed channels and this increases the self-jamming (own goal) probability $\rho(n)$ for the monitor.

\begin{figure}[!t]
\centering
\includegraphics[width=3.0in]{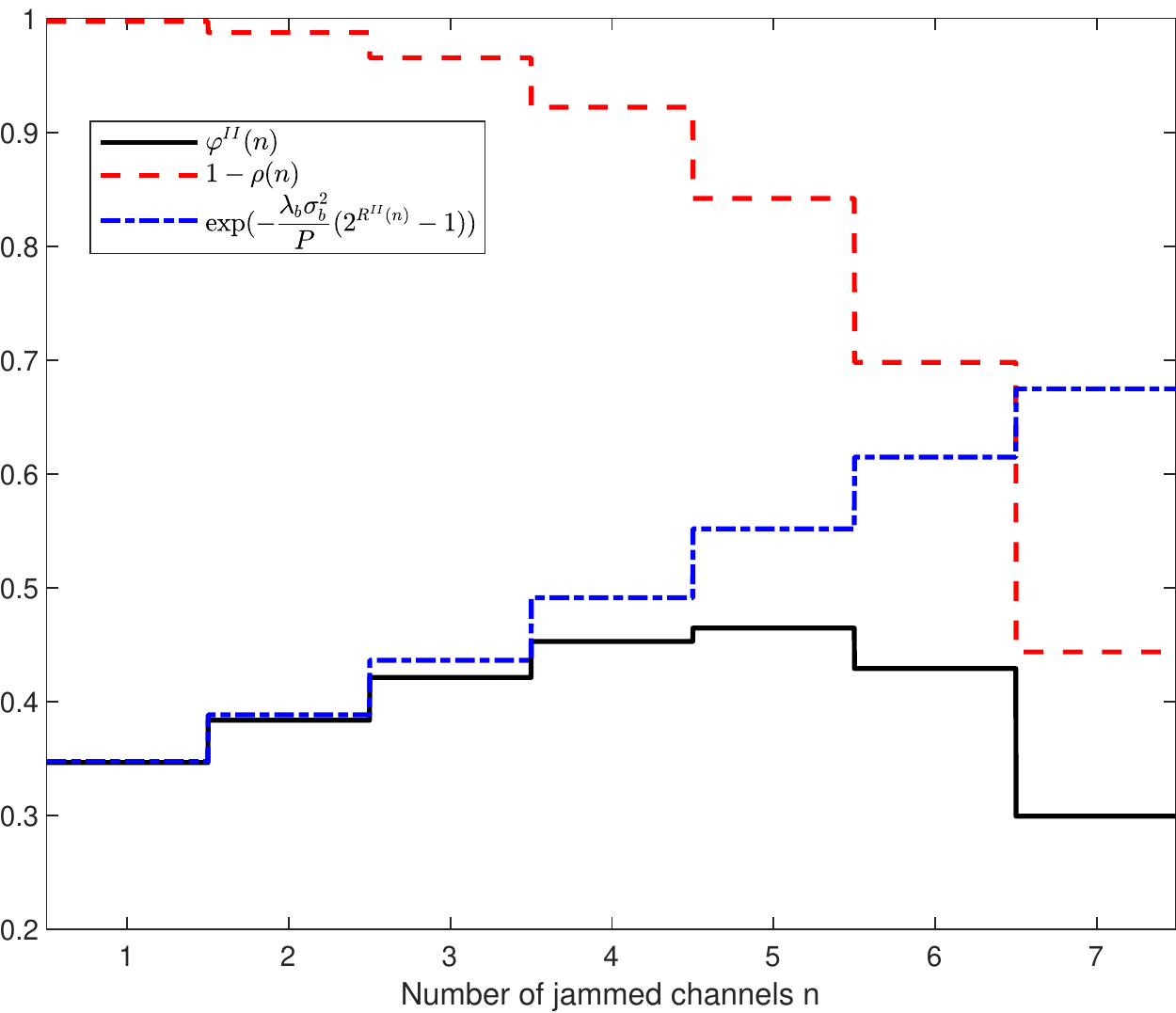}
\caption{$\varphi^{II}(n)$, $1-\rho(n)$ and non-outage probability at the monitor versus number of jammed channels $n$, where $N=8$, $\lambda_a=\lambda_b=1$, $\lambda_c=3$, $P=10$ dB, $Q_{\text{max}}=20$ dB, $\sigma_a^2=\sigma_b^2=1$ and $\delta=0.05$.}
\label{fig3}
\vspace{-0.8em}
\end{figure}

Next, we determine the non-outage probability $e^{-\lambda_b\sigma_b^2(2^{R^{II}(n)}-1)/P}$ at the monitor in \eqref{iv15}, which is a function of $R^{II}(n)$. Similar to $\rho(n)$, here $R^{II}(n)$ only depends on $n$.

\underline{\emph{Proposition 4.3:}} \ In the general multi-channel case, the suspicious link's transmission rate $R^{II}(n)$ in \eqref{iv15} is the unique solution to
\begin{equation}
\begin{aligned}
\Big(\!1\!-\!e^{-\frac{\lambda_a\sigma_a^2}{P}(2^{R^{II}}\!\!-1)}\!\Big)^{N-n} \! \bigg(\!1\!-\!\frac{\lambda_cPe^{-\frac{\lambda_a\sigma_a^2}{P}(2^{R^{II}}\!\!-1)}}{\lambda_cP\!+\!\lambda_a \frac{Q_{\max}}{n} (2^{R^{II}}\!\!-\!1)}\!\bigg)^{n}\!=\!\delta.
\label{iv17}
\end{aligned}
\end{equation}Further, $R^{II}(n)$ monotonically decreases as $n$ increases.

\begin{figure*}[ht]
\centering
\includegraphics[width=6.0in]{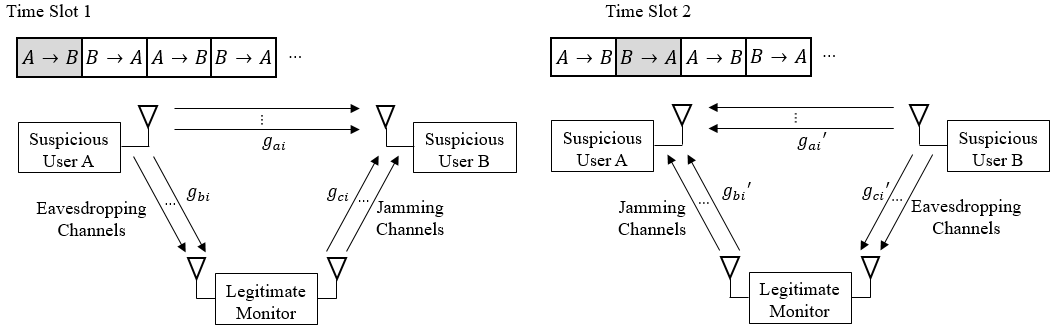}
\caption{Jamming-assisted eavesdropping over two-way communications (user A and user B alternately communicate with each other over different time slots).}
\label{fig4}
\vspace{-0.5em}
\end{figure*}

\begin{IEEEproof}
See Appendix D.
\end{IEEEproof}

The reason why $R^{II}(n)$ decreases with $n$ is because jamming more channels increases the chance that the ST chooses the jammed channels, and the ST will transmit at a lower rate to maintain target outage probability at the SR. A lower transmission rate $R^{II}(n)$ leads to a higher non-outage probability $e^{-\lambda_b\sigma_b^2(2^{R^{II}(n)}-1)/P}$ at the monitor, and thus the monitor can overhear more clearly.

Fig.~\ref{fig3} numerically illustrates $1-\rho(n)$ in \eqref{iv16}, non-outage probability in \eqref{iv17}, and their product $\varphi^{II}(n)$ in \eqref{iv15}. As $n$ increases, it is observed that $1-\rho(n)$ decreases and $e^{-\lambda_b\sigma_b^2(2^{R^{II}(n)}-1)/P}$ increases. To maximize $\varphi^{II}(n)$, there is thus a trade-off in deciding the optimal number of jammed channels $n^*$ (here $n^*=5$ in this numerical example).

It should be noted that it is still difficult to analytically solve $n^*$ in (P3) even by relaxing $n$ to be continuous for tractable analysis. This is because not only $R^{II}(n)$ in \eqref{iv17} is not in closed-form, but also it is difficult to approximate $\rho(n)$ in \eqref{iv16} to be a continuous function due to the combinatorial nature and the involved upper incomplete gamma function. Still, we can numerically obtain $n^*$, by a one-dimensional exhaustive search in the set $\{1, 2, \cdots, N-1\}$ with low computation complexity of $O(N)$.

\underline{\emph{Proposition 4.4:}} \ In jamming-assisted eavesdropping scheme, as the monitor's jamming power budget $Q_{\max}$ goes to zero, it is optimal for the monitor to jam as few channels as possible ($n^*=1$), and the eavesdropping success probability is
\begin{equation}
\begin{aligned}
\lim_{Q_{\max} \to 0^+}\varphi ^{II}(n^*=1)=\big(1-\frac{1}{N}\big)e^{\frac{\lambda_b\sigma_b^2}{\lambda_a\sigma_a^2}\ln\big(1-\delta^{\frac{1}{N}}\big)}.
\label{iv18}
\end{aligned}
\end{equation}On the other hand, as the monitor's jamming power budget $Q_{\max}$ goes to infinity, it is optimal for the monitor to maximally jam $n^*=N-1$ channels and eavesdrop the remaining one with ideal zero own goal probability, and the eavesdropping success probability is
\begin{equation}
\begin{aligned}
\lim_{Q_{\max} \to \infty}\varphi ^{II}(n^*=N-1)=e^{\frac{\lambda_b\sigma_b^2}{\lambda_a\sigma_a^2}\ln(1-\delta)}.
\label{iv19}
\end{aligned}
\end{equation}

\begin{IEEEproof}
When the monitor's jamming power budget goes to zero, the jammed channels and unjammed channels are the same in distribution, while the half-duplex monitor cannot eavesdrop the jammed channels, making the own goal probability proportional to $n$, i.e.,
\begin{equation}
\begin{aligned}
\lim_{Q_{\max} \to 0^+}\rho(n) = \frac{n}{N}. \nonumber
\end{aligned}
\end{equation}Meanwhile, the transmission rate $R^{II}(n)$ in \eqref{iv17} now becomes the same as constant transmission rate of passive eavesdropping in \eqref{ii5}. Thus, the eavesdropping success probability in \eqref{iv15} becomes
\begin{equation}
\begin{aligned}
\lim_{Q_{\max} \to 0^+}\varphi ^{II}(n)= \big(1-\frac{n}{N}\big)e^{\frac{\lambda_b\sigma_b^2}{\lambda_a\sigma_a^2}\ln\big(1-\delta^{\frac{1}{N}}\big)}.
\label{iv20}
\end{aligned}
\end{equation}As we can see from \eqref{iv20}, $\lim_{Q_{\max} \to 0^+}\varphi ^{II}(n)$ monotonically decreases as $n$ increases, thus it is optimal to only jam $n^*=1$ channel when employing the jamming-assisted eavesdropping.

On the other hand, as jamming power budget goes to infinity, the own goal probability becomes $\lim_{Q_{\max} \to \infty}\rho(n)=0$, as the severely jammed channel will never be chosen by the ST. Meanwhile, the transmission rate $R^{II}(n)$ becomes
\begin{equation}
\begin{aligned}
\lim_{Q_{\max} \to \infty}R^{II}(n)= \log_2\bigg(1+\frac{P}{\lambda_a\sigma_a^2}\ln\frac{1}{1-\delta^{\frac{1}{N-n}}}\bigg). \nonumber
\end{aligned}
\end{equation}Thus, the eavesdropping success probability becomes
\begin{equation}
\begin{aligned}
\lim_{Q_{\max} \to \infty}\varphi ^{II}(n)= e^{\frac{\lambda_b\sigma_b^2}{\lambda_a\sigma_a^2}\ln\big(1-\delta^{\frac{1}{N-n}}\big)}.
\label{iv21}
\end{aligned}
\end{equation}As we can see from (21), $\lim_{Q_{\max} \to \infty}\varphi ^{II}(n)$ monotonically increases as $n$ increases, thus it is optimal to jam $n^*=N-1$ channels in this case.
\end{IEEEproof}

Note that under optimal $n^*$, if $\varphi ^{II}(n^*)>\varphi ^{I}$, the legitimate monitor will perform jamming-assisted eavesdropping, otherwise it will perform passive eavesdropping.

\section{Extension to eavesdropping two-way communications}

So far, we have considered the one-way communication from the ST to the SR for the suspicious link, while in practice, the two users may need to alternately exchange information with each other over time periods or fading blocks (see Fig.~\ref{fig4}). Our jamming-assisted eavesdropping approach in Section IV is designed for the one-way communication, and will be extended in this section for the two-way communications.

The distributions of channel power gains of the communication from user A to user B follow the same model defined in Section II, i.e., $g_{ai}\sim \exp(\lambda_a)$, $g_{bi}\sim \exp(\lambda_b)$, $g_{ci}\sim \exp(\lambda_c)$. Due to the reciprocity of wireless channel, the suspicious communication channels are the same from user B to user A, i.e., $g_{ai}'\sim \exp(\lambda_a)$. However, the original jamming channels now become eavesdropping channels, i.e., $g_{bi}'\sim \exp(\lambda_c)$, and the original eavesdropping channels now become jamming channels, i.e., $g_{ci}'\sim \exp(\lambda_b)$. As we can see, the optimal number of jammed channels are in general different for user A to B communication ($n_{AB}^*$ as computed in last section) and user B to A communication ($n_{BA}^*$), where subscript $(\cdot)_{AB}$ denotes the communication direction from user A to user B, and $(\cdot)_{BA}$\ denotes the communication from user B to user A. But alternately jamming $n_{AB}^*\neq n_{BA}^*$ channels over time will easily arouse the suspicion of suspicious users by examining the channel statistics. Then the suspicious link can tell that the monitor is a jammer instead of a normal D2D user with time-division-duplex.

To intercept the two-way communications between user A and user B, we need to balance these two communication ways for maximizing the minimum eavesdropping success probability between both communication ways, keep the same jammed channels and jamming power in the long run. Without loss of generality, if $n$ out of $N$ i.i.d. fading channels are jammed in two-way communications, we consider the monitor picks the first $n$ out of $N$ channels to jam. Similar to Proposition 4.1 in the one-way communication, the monitor should also use up all the jamming power budget and evenly partition over the jammed channels in two-way communications. Thus, the max-min optimization problem can be reformulated as follows.\footnote{Our problem (P4) can further include different weights for the two communication ways. For example, if the monitor values the message from user A (e.g., a leader of a criminal gang) to user B (e.g., a follower) more important, it will assign a large weight to this way's eavesdropping success probability.}
\begin{equation}
\begin{aligned}
(\mathrm{P4}): &\max_n\max\{\min\{\varphi _{AB}^{I},\varphi _{BA}^{I}\},\min\{\varphi _{A B}^{II}(n),\varphi _{BA}^{II}(n)\}\} \\
&\ \mathrm{s.t.} \ 1\leq n\leq N-1, n\in\mathbb{Z}.\nonumber
\label{p4}
\end{aligned}
\end{equation}

(P4) compares the performance of jamming-assisted eavesdropping $\min\{\varphi _{AB}^{II}(n),\varphi _{BA}^{II}(n)\}$ and the performance of passive eavesdropping $\min\{\varphi _{AB}^{I},\varphi _{BA}^{I}\}$ in two-way communications. Since both $\varphi _{AB}^{I}$ and $\varphi _{BA}^{I}$ are constants, we can only focus on optimizing the performance of jamming-assisted eavesdropping $\min\{\varphi _{AB}^{II}(n),\varphi _{BA}^{II}(n)\}$. To numerically solve this integer programming, we can obtain the optimal number of jammed channels $n^*$ in two-way communications, by efficiently performing one-dimensional exhaustive search in the set $\{1, 2, \cdots, N-1\}$ with low computation complexity of $O(N)$. Then we compare $\min\{\varphi _{AB}^{I},\varphi _{BA}^{I}\}$ with $\min\{\varphi _{AB}^{II}(n^*),\varphi _{BA}^{II}(n^*)\}$: if the former is smaller, the monitor will perform jamming-assisted eavesdropping with $n^*$ jammed channels, otherwise the monitor will perform passive eavesdropping.

We first use a numerical example to illustrate the eavesdropping success probability of jamming-assisted eavesdropping in two-way communications. Assuming there are $N=8$ parallel channels, we set the mean channel power gains to be $1/\lambda_a=1/5,\ 1/\lambda_b=1$ and $1/\lambda_c=1/4$, additive white Gaussian noise (AWGN) power as $\sigma_a^2=\sigma_b^2=\sigma_c^2=1$, target outage probability at user A/B as $\delta=0.05$, transmitting power of user A/B as $P=10$ dB and jamming power budget as $Q_{\max}=20$ dB.

\begin{figure}[!t]
\centering
\includegraphics[width=3.0in]{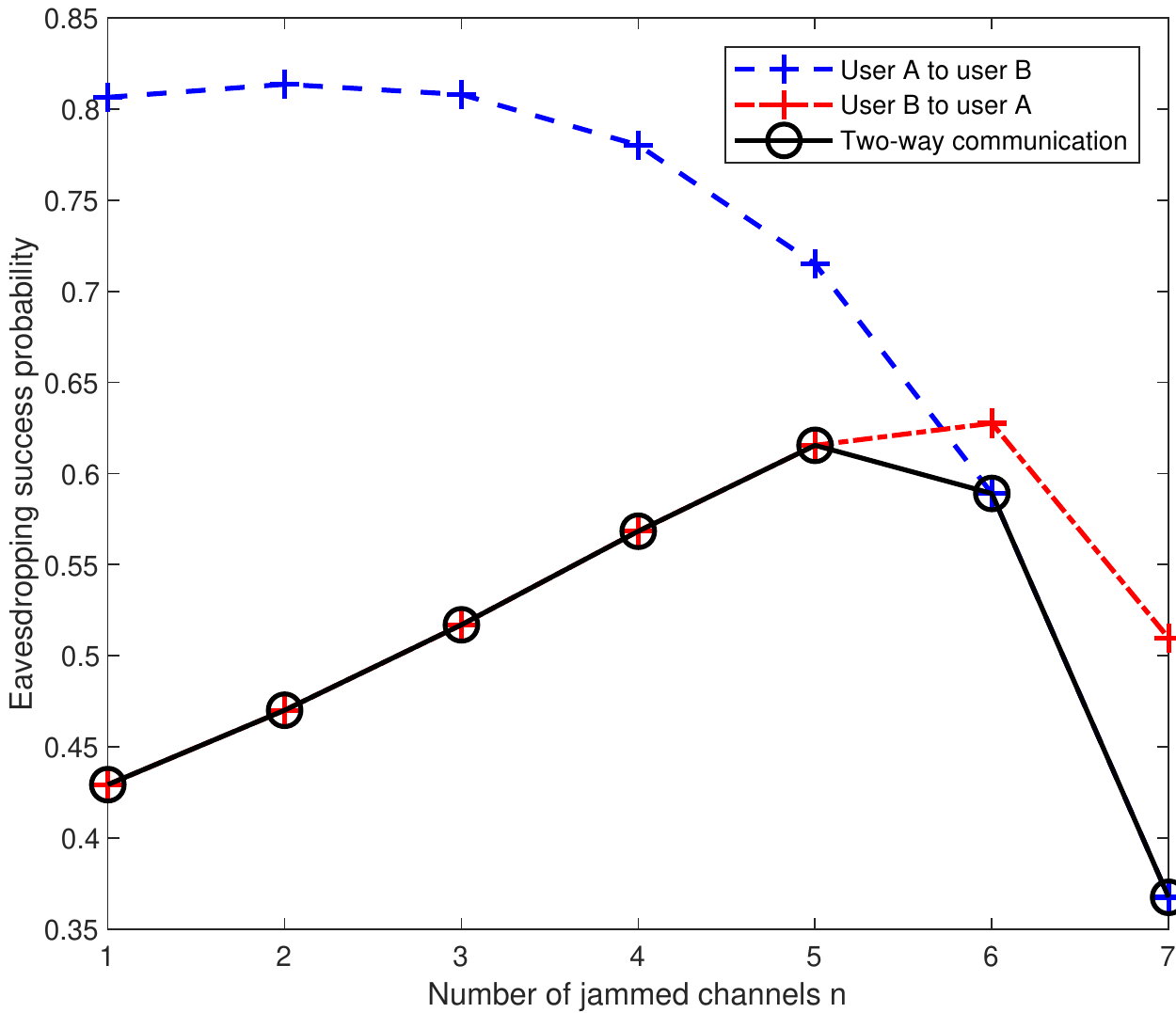}
\caption{Eavesdropping success probability in jamming-assisted eavesdropping of two-way communications, which is the minimum of the eavesdropping success probabilities from user A to B and from user B to A.}
\label{fig5}
\vspace{-0.8em}
\end{figure}

As Fig.~\ref{fig5} shows, the optimal number of jammed channels $n_{AB}^*$ for suspicious user A's communication to user B is $2$, while the optimal number of jammed channels $n_{BA}^*$ for suspicious user B's communication to user A is $6$; and we find the optimal number of jammed channels $n^* = 5$ to balance between the communications from user A to B and from user B to A.

\begin{figure*}[ht]
\begin{minipage}[t]{0.47\textwidth}
\centering
\includegraphics[width=3.0in]{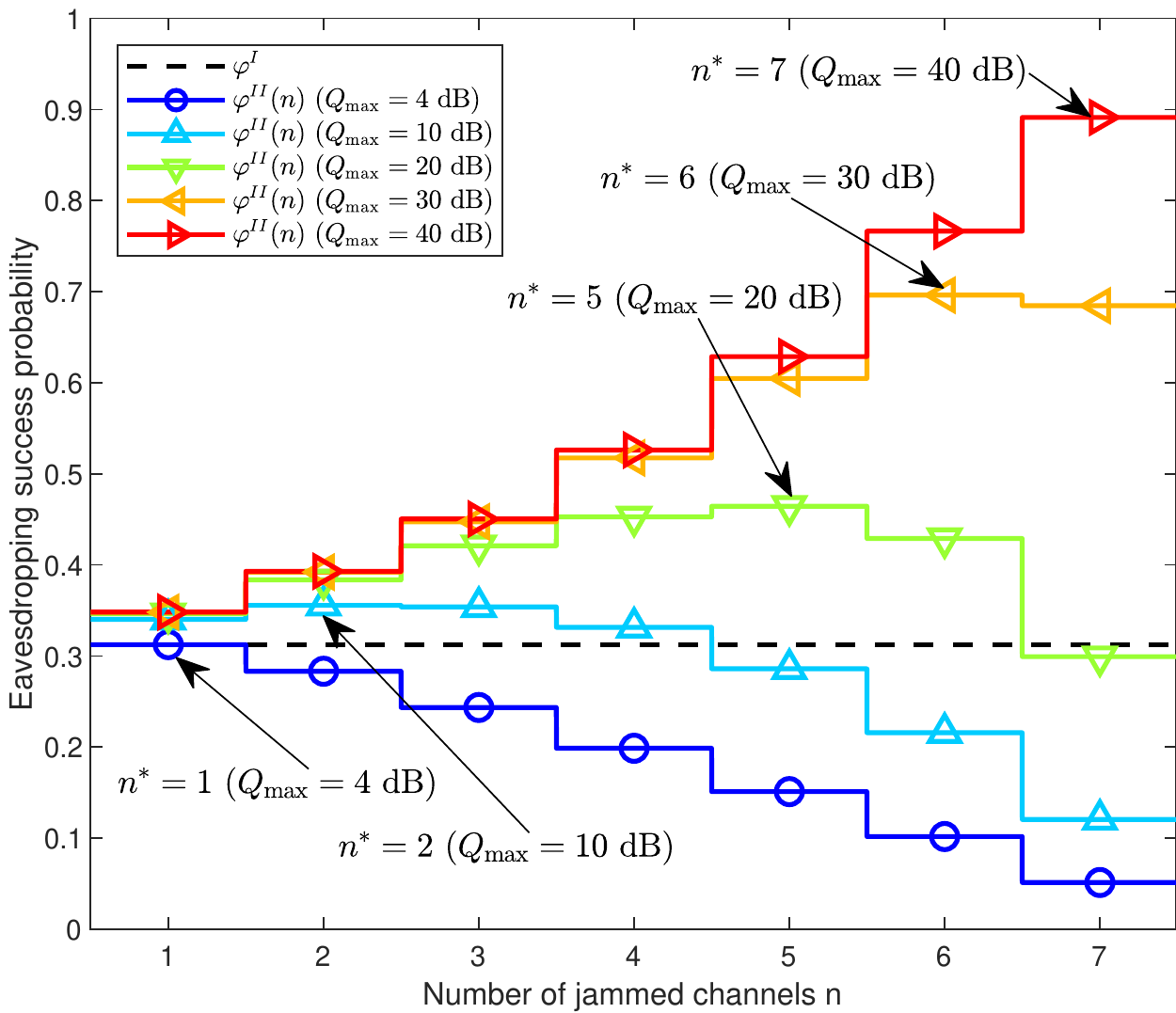}
\caption{Eavesdropping success probability $\varphi^{II}$ versus number of jammed channels $n$ under different power budget $Q_{\max}$'s.}
\label{fig6}
\end{minipage}\hfill
\begin{minipage}[t]{0.47\textwidth}
\centering
\includegraphics[width=3.0in]{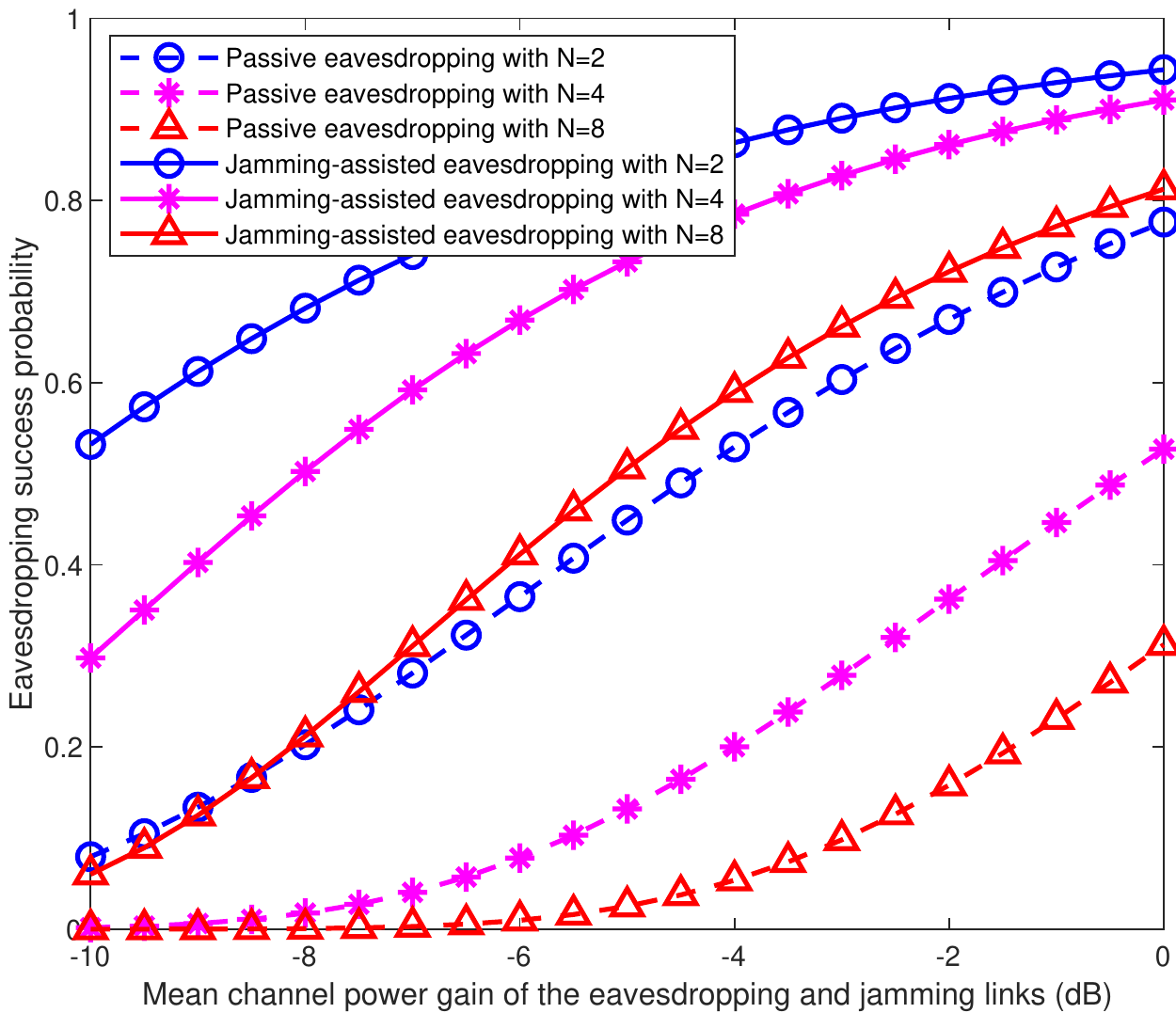}
\caption{Eavesdropping success probability versus mean channel power gain of the eavesdropping and jamming links with different total numbers of channels $N$'s.}
\label{fig7}
\end{minipage}
\end{figure*}

It is difficult to analytically derive the optimal solution to (P4), due to the non-concave objective involving incomplete gamma function and the discrete nature of decision variable $n$. Despite of these, we still manage to derive some analytical results by assuming the one-way eavesdropping success probability $\varphi^{II}(n)$ (i.e., $\varphi_{AB}^{II}(n)$ for user A to B and $\varphi_{BA}^{II}(n)$ for user B to A) is unimodal (having only one peak) or monotonic in $n$ as in Fig.~\ref{fig3}. Actually, this is always the case in our extensive simulations though it is difficult to rigorously prove.

\underline{\emph{Proposition 5.1:}} \ Assuming the objective functions $\varphi_{AB}^{II}(n)$ and $\varphi_{BA}^{II}(n)$ in (P4) are unimodal or monotonic in $n$, the optimal jamming-assisted eavesdropping scheme is given as follows, depending on the jamming power budget $Q_{\max}$.

\begin{itemize}
\item If the monitor's jamming power budget $Q_{\max}$ is low (i.e., $Q_{\max}<\underline{Q}$), jamming more channels hurts the eavesdropping performances on both ways, where $\underline{Q}=\min\{Q_{\max}|\varphi_{AB}^{II}(1)\geq\varphi_{AB}^{II}(2),\varphi_{BA}^{II}(1)\geq\varphi_{BA}^{II}(2)\}$. In this case, it is optimal for the monitor to minimally jam $n^*=1$ channel. As a special case, when $Q_{\max}$ goes to zero, $n^*=1$ (as a two-way extension of Proposition 4.4).
\item If the monitor's jamming power budget $Q_{\max}$ is high (i.e., $Q_{\max}>\bar{Q}$), jamming more channels improves the eavesdropping performances on both ways, where $\bar{Q}=\max\{Q_{\max}|\varphi_{AB}^{II}(N-2)\leq\varphi_{AB}^{II}(N-1),\varphi_{BA}^{II}(N-2)\leq\varphi_{BA}^{II}(N-1)\}$. In this case, it is optimal for the monitor to maximally jam $n^*=N-1$ channels. As a special case, when $Q_{\max}$ goes to infinity, $n^*=N-1$ (as a two-way extension of Proposition 4.4).
\item If the jamming power budget $Q_{\max}$ is medium (i.e., $\underline{Q}\leq Q_{\max}\leq \bar{Q}$), the optimal number of jammed channels $n^*$ is between $n_{AB}^*$ and $n_{BA}^*$, by balancing the eavesdropping performances of the two ways.
\end{itemize}

\begin{IEEEproof}
See Appendix E.
\end{IEEEproof}

\section{Numerical Results}

In this section, we provide more numerical results to validate our studies and designs. Assuming there are $N=8$ parallel channels, we set the mean channel power gain of the suspicious communication link, eavesdropping link and jamming link in the one-way communication to be $1/\lambda_a=1,\ 1/\lambda_b=1$ and $1/\lambda_c=1/3$, respectively. We also set the AWGN power as $\sigma_a^2=\sigma_b^2=\sigma_c^2=1$, the target outage probability at the SR as $\delta=0.05$ and transmitting power of the ST as $P=10$ dB.

Fig.~\ref{fig6} shows the eavesdropping success probability $\varphi^{II}$ as a function of $n$ and $Q_{\max}$ for the case of one-way communication from suspicious user A to user B. When $Q_{\max}$ is small (e.g., $Q_{\max}=4$ dB curve in Fig.~\ref{fig6}), the monitor will perform passive eavesdropping. When $Q_{\max}$ is sufficiently large (starting from $Q_{\max}=10$ dB), the monitor will jam to assist eavesdropping for a larger eavesdropping success probability. We can see that as $Q_{\max}$ increases, the monitor will jam more channels by optimally controlling the trade-off between own goal probability and transmission rate. When $Q_{\max}$ is further large (e.g., $Q_{\max}=40$ dB), the monitor will optimally jam $N-1$ channels to overhear the lowest-rate suspicious communication in the remaining single channel, without worrying about the own goal of self-jamming. This result is consistent with Proposition 4.4.

Then we consider the passive eavesdropping as a benchmark for performance comparison. We set the mean channel power gain of the suspicious communication link to be $1/\lambda_a=1$ and jamming power budget to be $Q_{\max}=30$ dB. Here we consider the monitor is far away from the ST and SR, thus the mean channel power gains of eavesdropping channels and jamming channels are nearly the same, i.e., $1/\lambda_b=1/\lambda_c$.

Fig.~\ref{fig7} shows the eavesdropping success probability as a function of $N$ and mean channel power gain of eavesdropping and jamming links. As their mean channel power gains increase, eavesdropping success probabilities of both jamming-assisted and passive eavesdropping increase. But jamming-assisted eavesdropping greatly outperforms passive eavesdropping. The performance of jamming-assisted eavesdropping is better when total number of channels is smaller (e.g., $N=2$ in Fig.~\ref{fig7}), as the monitor can more efficiently induce the suspicious link to use a smaller subset of unjammed channels and a lower transmission rate under the same jamming power budget. While the performance gain of jamming-assisted eavesdropping comparing with passive eavesdropping is greater when total number of channels is large (e.g., $N=8$ in Fig.~\ref{fig7}), as more channels provides more degrees of freedom for jamming.

\begin{figure*}[ht]
\centering
\subfigure[Optimal number of jammed channels $n^*$ versus different monitor's locations in the 2D ground plane ($N=8$ channels in total).]{
\includegraphics[width=2.2in]{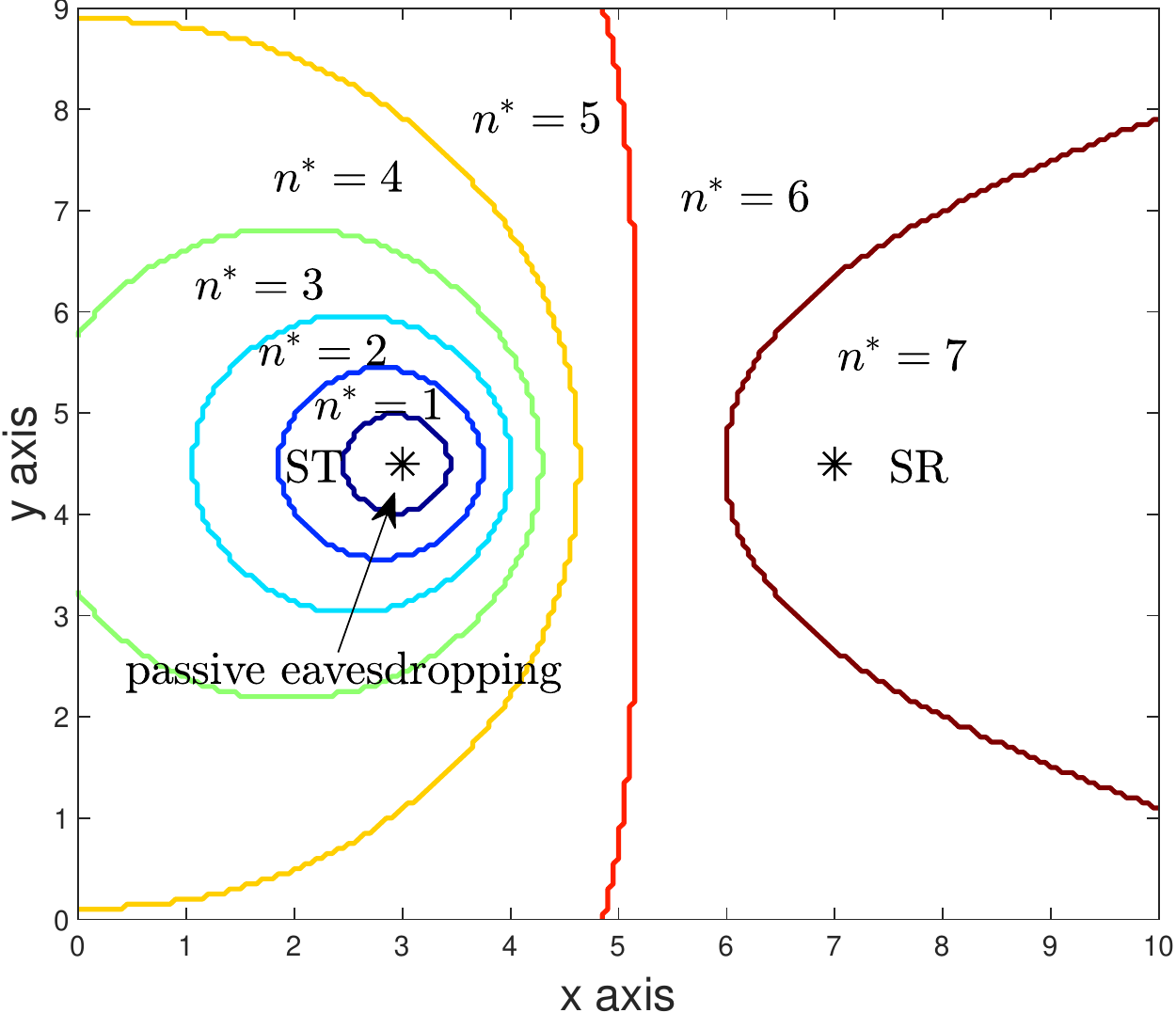}}
\hspace{0.1in}
\subfigure[Eavesdropping success probability of jamming-assisted eavesdropping versus different monitor's locations in the 2D ground plane.]{
\includegraphics[width=2.2in]{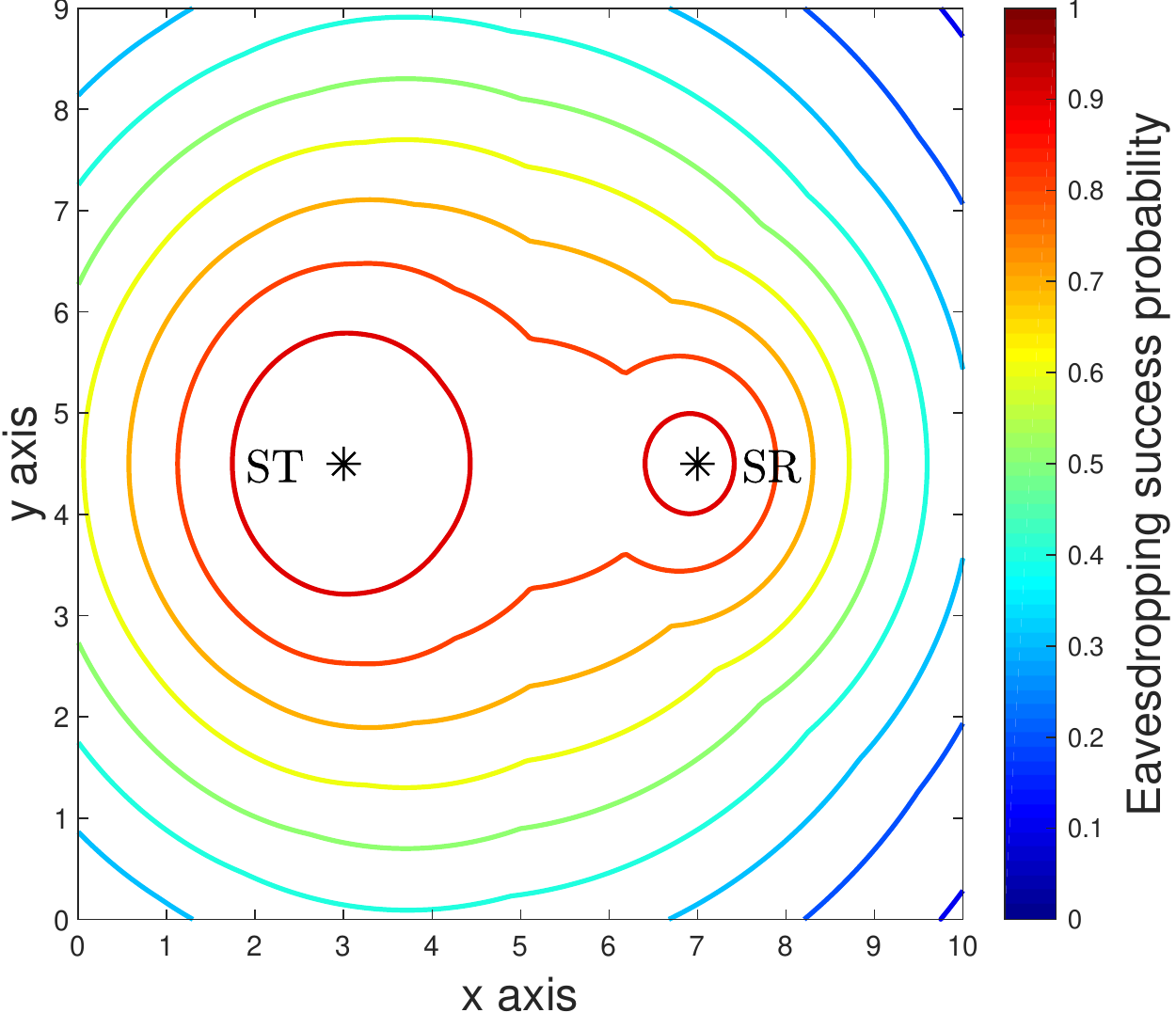}}
\hspace{0.1in}
\subfigure[Eavesdropping success probability of passive eavesdropping versus different monitor's locations in the 2D ground plane.]{
\includegraphics[width=2.2in]{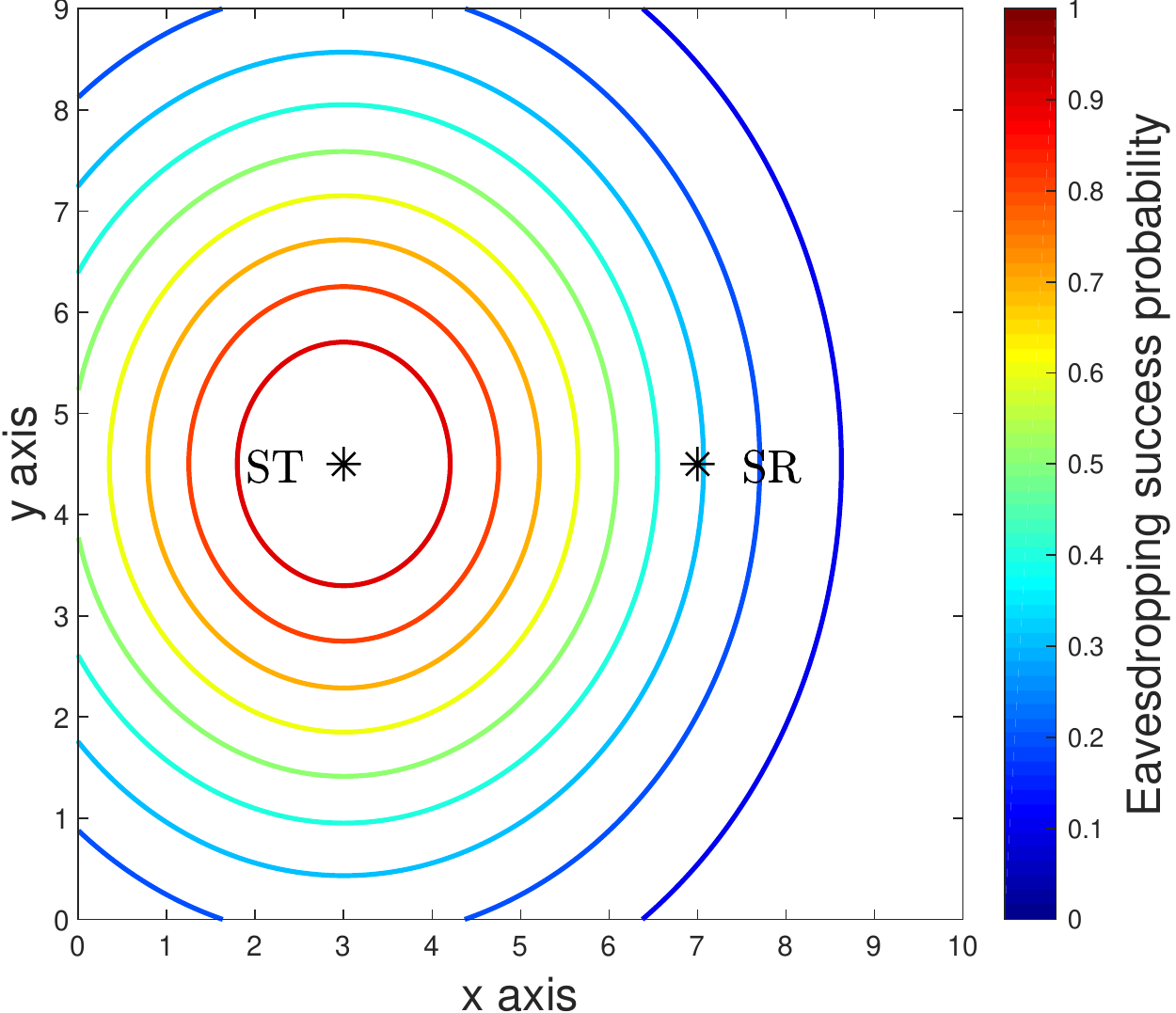}}
\caption{Effect of monitor's geometric locations on jamming strategy design, jamming-assisted eavesdropping performance, and passive eavesdropping performance.}
\label{fig8}
\end{figure*}
\begin{figure*}[ht]
\centering
\subfigure[]{
\includegraphics[width=2.9in]{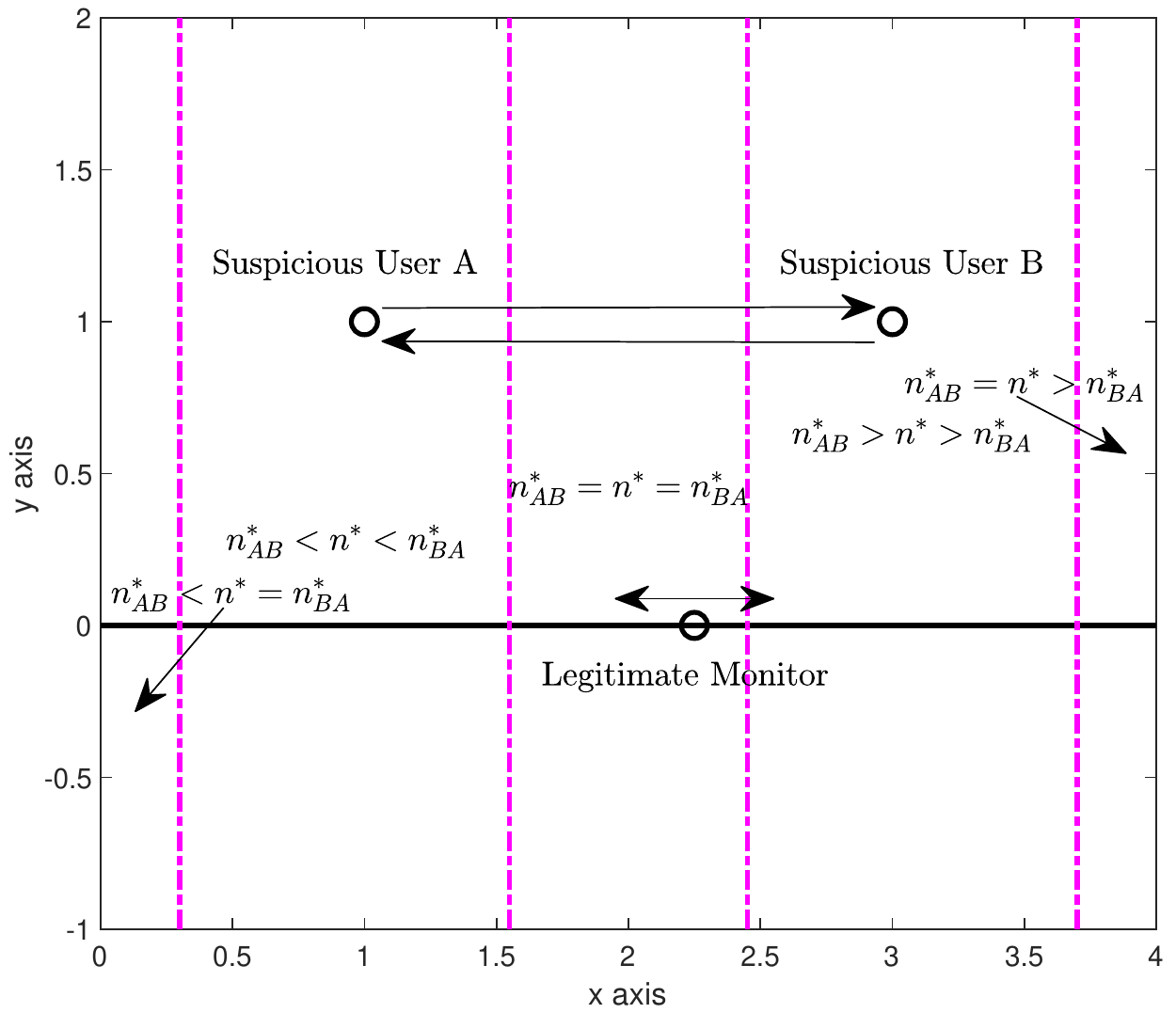}}
\hspace{0.5in}
\subfigure[]{
\includegraphics[width=2.9in]{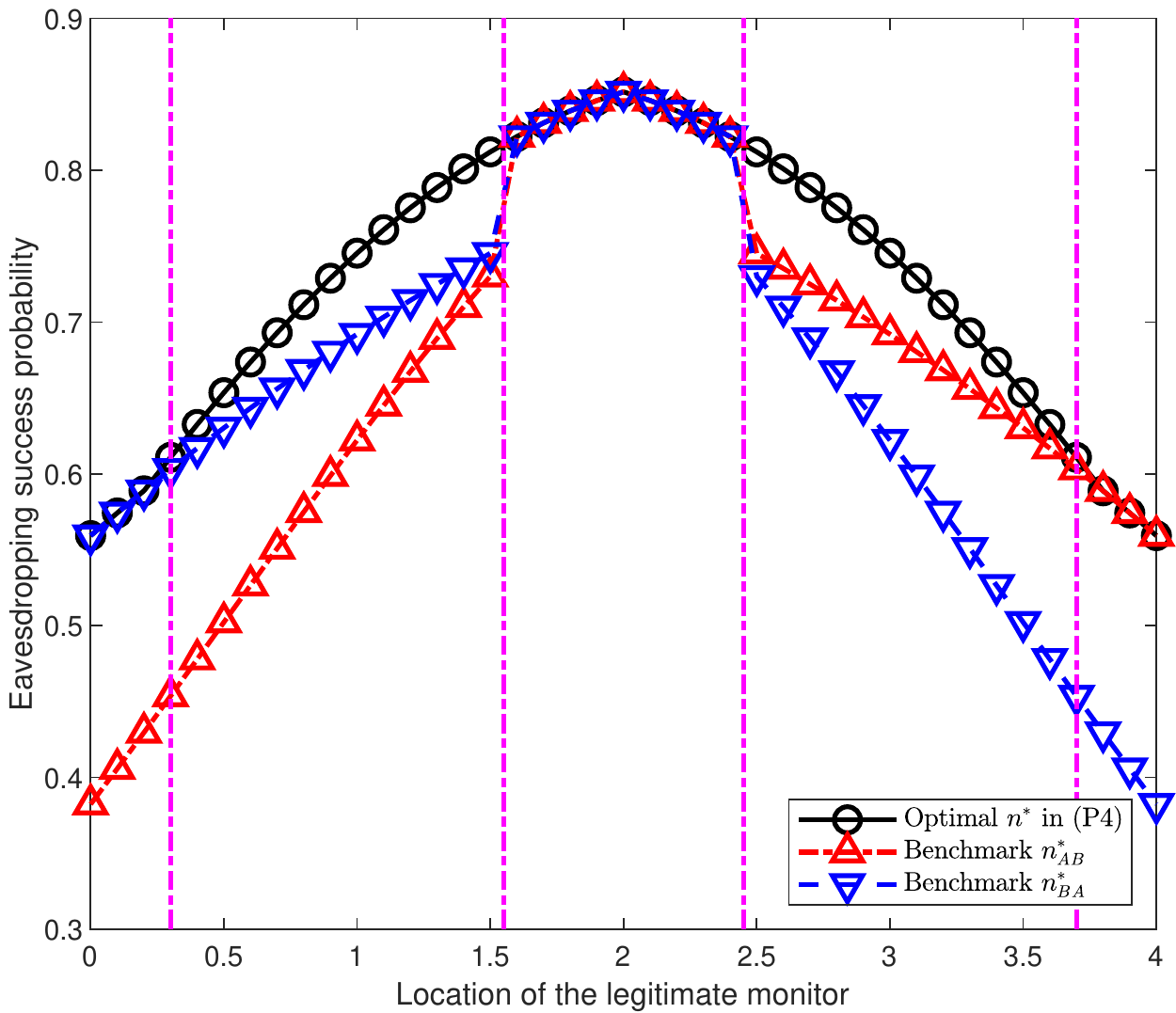}}
\caption{Eavesdropping success probability versus monitor's location in two-way communications.}
\label{fig9}
\end{figure*}

Next, we study the effect of the monitor's location. To capture the effect of large-scale fading, we consider that for any two points with coordinates $(x_1,y_1)$ and $(x_2,y_2)$ in the two-dimensional (2D) ground plane, the mean channel power gain between the two points is inversely proportional to the square of their distance, i.e., $1/\lambda=1/\big((x_1-x_2)^2+(y_1-y_2)^2\big)$. The ST is located at $(3,4.5)$, and the SR is located at $(7,4.5)$. The distances are normalized with transmit power. The legitimate monitor is placed in different locations in this plane, and its jamming power budget is $Q_{\max}=30$ dB.

Fig.~\ref{fig8}(a) shows the optimal number of jammed channels $n^*$ in different monitor's locations in the 2D ground plane. When the monitor is close to the ST, passive eavesdropping can already provide good eavesdropping performance. When the monitor is moving away from the ST, passive eavesdropping can no longer provide good eavesdropping performance, and the monitor will jam more channels to lower the suspicious link's transmission rate in order to overhear more clearly. When the monitor is close to the SR, which means now the efficiency of jamming is high, the monitor will jam most channels (up to $N-1$) for improving eavesdropping performance.

Fig.~\ref{fig8}(b) and Fig.~\ref{fig8}(c) compare the eavesdropping success probability of jamming-assisted eavesdropping and passive eavesdropping with different monitor's locations in the 2D ground plane. As we can see, the eavesdropping success probability of passive eavesdropping is fully determined by the distance between the monitor and the ST, and a good eavesdropping performance can only be guaranteed when the monitor is close to the ST. While our proposed jamming-assisted eavesdropping greatly outperforms passive eavesdropping even when the monitor is not close to the ST, because it can efficiently jam the SR and drive the ST to transmit in a smaller subset of channels with a lower transmission rate so that the monitor can eavesdrop more effectively. The performance gain of jamming-assisted eavesdropping is significant when the monitor is close to the SR, since now the efficiency of jamming is high. This clearly shows that passive eavesdropping is dramatically sensitive to the ST-monitor distance, while our proposed jamming-assisted eavesdropping is no longer that sensitive even when the monitor is geometrically far away from the ST.

Finally, we examine the performance of jamming-assisted eavesdropping in two-way communications. Here we still assume the mean channel power gain between any two points is inversely proportional to the square of their distance, similar to the previous simulation. Suspicious user A is located at $(1,1)$ and suspicious user B at $(3,1)$. The legitimate monitor can be at any point between $(0,0)$ and $(4,0)$ to eavesdrop the two-way communications, and its jamming power budget is set to $Q_{\max}=30$ dB. We provide two benchmark cases for performance comparison with our optimal solution. Benchmark $n_{AB}^*$ tells that the monitor just focuses on the one-way communication from A to B, and jams $n_{AB}^*$ channels according to the one-way surveillance problem in (P3). Meanwhile, benchmark $n_{BA}^*$ focuses on the one-way communication from B to A, and jams $n_{BA}^*$ channels.

Fig.~\ref{fig9}(a) compares the optimal number of jammed channels $n^*$ in (P4) with $n_{AB}^*$ and $n_{BA}^*$ when the monitor moves horizontally between $(0,0)$ and $(4,0)$. Fig.~\ref{fig9}(b) shows the eavesdropping success probability versus the monitor's location in two-way communications. We have the following observations.

\begin{itemize}
\item{When the monitor is between $(0,0)$ and $(0.3,0)$, it is far away from user B and the overall eavesdropping performance is bottlenecked by user B as the ST, thus jamming $n_{BA}^*$ channels will give the best eavesdropping performance as the optimal $n^*$. Note that $n_{AB}^*$ is smaller than $n_{BA}^*$ ($n^*$) in this case, since user A is much closer to the monitor than user B, which is consistent with Fig.~\ref{fig8}(a) in the sense that the monitor will jam more channels when it is moving away from the ST.}
\item{When the monitor is moving from $(0.3,0)$ to $(1.6,0)$, it is getting closer to user A than user B, thus the eavesdropping performance of benchmark $n_{AB}^*$ improves more significantly than that of benchmark $n_{BA}^*$. Still, $n^*$ outperforms benchmarks $n_{AB}^*$ and $n_{BA}^*$. Note that $n_{BA}^*$ is greater than $n^*$ in this case, because now the monitor is very close to user A, and according to Fig.~\ref{fig8}(a), the monitor will jam most channels when it is close to the SR.}
\item{When the monitor is between $(1.6,0)$ and $(2,0)$, the monitor's distances to users A and B are close, thus both benchmarks $n_{AB}^*$ and $n_{BA}^*$ are symmetric and achieve the optimal eavesdropping performance as $n^*$. Note that in Fig.~\ref{fig9}(b) the sharp performance increase of benchmarks $n_{AB}^*$ and $n_{BA}^*$ at point $(1.6,0)$ is caused by changing the number of jammed channels. Taking $n_{AB}^*$ as an example, the monitor is moving closer to user B (SR in $n_{AB}^*$'s point of view), and at the point $(1.6,0)$, the monitor increases the number of jammed channels due to better jamming efficiency, which causes the obvious eavesdropping performance change.}
\end{itemize}

Finally, the eavesdropping success probability reaches the maximum when the monitor is at $(2,0)$, since now the monitor's distances to user A and user B are the same. The eavesdropping performance analysis of the monitor moving from $(2,0)$ to $(4,0)$ is similar as above by symmetry.

\section{Conclusions}

This paper proposes a new wireless security model, which is jamming-assisted legitimate eavesdropping over parallel independently fading channels. The legitimate monitor uses jamming in order to achieve better eavesdropping performance. Assuming Rayleigh fading, we formulate the optimization problem for jamming design as a mixed integer nonlinear programming (MINLP). Despite its non-convexity, we show that the legitimate monitor should use the maximum jamming power for the best eavesdropping performance if it decides to jam. Then we simplify the MINLP to integer programming and further show that there is a trade-off in deciding the number of jammed channels in the general multi-channel case, where jamming more channels helps reduce the suspicious communication rate for overhearing more clearly, but at the risk that the ST is more likely to choose jammed channels to transmit and as a result cannot be overheard. Finally, we extend our study to two-way communications, and show another trade-off in deciding the common jammed channels for balancing the bidirectional eavesdropping performances. Numerical results show that our jamming-assisted eavesdropping schemes greatly improve eavesdropping success probability comparing with conventional passive eavesdropping.

This work can be extended in various directions. For example, the suspicious link can transmit at multiple channels and perform combining at the receiver, or there can be multiple suspicious link pairs, which will bring more challenges to legitimate eavesdropping. The more general case of parallel channels with correlated (non-independent) fading is also worth investigating in future work.

\section*{Appendix}

\subsection{Proof of Lemma 2.1}

We aim to show the distribution of SINR $Y=\frac{g_{ai}P}{g_{ci}Q_i+\sigma_a^2}$ on jammed channel $i$. Denote its numerator and denominator as $Y_1=g_{ai}P$ and $Y_2=g_{ci}Q_i+\sigma_a^2$, respectively. As $g_{ai}$ and $g_{ci}$ follow independent exponential distributions with mean $1/\lambda_a$ and $1/\lambda_c$, respectively, we have
\begin{equation}
\begin{aligned}
f_{Y_1}(y_1) = \frac{\lambda_a}{P}e^{-\frac{\lambda_a}{P}y_1}, \ \ \ \ \ \ \ \ \ \ \ \ \  y_1\geq0, \nonumber
\end{aligned}
\end{equation}
\begin{equation}
\begin{aligned}
f_{Y_2}(y_2) = \frac{\lambda_c}{Q_i}e^{-\frac{\lambda_c}{Q_i}(y_2-\sigma_a^2)}, \ \ \ \ \ \ \ \ y_2\geq\sigma_a^2. \nonumber
\end{aligned}
\end{equation}

The probability density function (PDF) of $Y=Y_1/Y_2$ can be calculated as follows
\begin{equation}
\begin{aligned}
f_{Y}(y) = \frac{\lambda_a\lambda_c}{PQ_i}e^{\frac{\lambda_c\sigma_a^2}{Q_i}}\int_{\sigma_a^2}^{\infty} y_2e^{-(\frac{\lambda_c}{Q_i}+\frac{\lambda_ay}{P})y_2} \,dy_2. \nonumber
\end{aligned}
\end{equation}

With the help of [25, Eq. $3.351.2$], we have
\begin{equation}
\begin{aligned}
f_{Y}(y)=\frac{\lambda_a\lambda_c(\lambda_cP\sigma_a^2+PQ_i+\lambda_aQ_i\sigma_a^2y)}{(\lambda_cP+\lambda_aQ_iy)^2}e^{-\frac{\lambda_a\sigma_a^2y}{P}}, \ \  y\geq0, \nonumber
\end{aligned}
\end{equation}and the CDF of $Y$ can be calculated as
\begin{equation}
\begin{aligned}
\int_{0}^{y}f_{Y}(y) \,dy = 1-\frac{\lambda_cPe^{-\frac{\lambda_a\sigma_a^2}{P}y}}{\lambda_cP+\lambda_a Q_i y}, \ \ \ \  y\geq 0. \nonumber
\end{aligned}
\end{equation}

\subsection{Proof of Proposition 4.1}

For certain jamming scheme $S_1$, the own goal probability can be expressed as
\begin{equation}
\begin{aligned}
\rho_{S_1}(n)=&\mathbb{P}\bigg(\max\Big\{\frac{g_{a1}P}{g_{c1}Q_1+\sigma_a^2},\cdots,\frac{g_{aj}P}{g_{cj}Q_j+\sigma_a^2},\cdots, \\ &\frac{g_{an}P}{g_{cn}Q_n+\sigma_a^2}\Big\}>\max\Big\{\frac{g_{a1}P}{\sigma_a^2},\cdots, \frac{g_{a(N-n)}P}{\sigma_a^2}\Big\}\bigg).\nonumber
\end{aligned}
\end{equation}

There exists a jamming scheme $S_2$, where the monitor reduces jamming power $Q_j$ on channel $j$ to $Q_j-\Delta$ with $\Delta>0$, for which
\begin{equation}
\begin{aligned}
\rho_{S_2}(n)=\mathbb{P}\bigg(\max\Big\{\frac{g_{a1}P}{g_{c1}Q_1\!+\!\sigma_a^2},\cdots,\frac{g_{aj}P}{g_{cj}(Q_j\!-\!\Delta)\!+\!\sigma_a^2},\cdots, \\ \frac{g_{an}P}{g_{cn}Q_n\!+\!\sigma_a^2}\Big\}>\max\Big\{\frac{g_{a1}P}{\sigma_a^2},\cdots, \frac{g_{a(N-n)}P}{\sigma_a^2}\Big\}\bigg).\nonumber
\end{aligned}
\end{equation}

Clearly, $\rho_{S_2}(n)$ is larger than $\rho_{S_1}(n)$.\\

For certain jamming scheme $S_1$, the transmission rate of the ST satisfies
\begin{equation}
\begin{aligned}
\!\Big(\!1\!-\!e^{-\frac{\lambda_a\sigma_a^2}{P}(2^{R_{S_1}^{II}}-1)}\!\Big)^{\!N-n} \!\prod\limits_{i=1}^n\!\bigg(\!1\!-\!\frac{\lambda_cPe^{-\frac{\lambda_a\sigma_a^2}{P}(2^{R_{S_1}^{II}}-1)}}{\lambda_cP\!+\!\lambda_a Q_i (2^{R_{S_1}^{II}}\!-\!1)}\!\bigg)\!\!=\!\delta.\nonumber
\end{aligned}
\end{equation}

There exists a jamming scheme $S_2$, where the monitor reduces jamming power $Q_j$ on channel $j$ to $Q_j-\Delta$, for which
\begin{equation}
\begin{aligned}
\Big(\!1\!-\!e^{-\frac{\lambda_a\sigma_a^2}{P}(2^{R_{S_2}^{II}}-1)}\!\Big)^{\!N-n}\!\bigg(\!1\!-\!\frac{\lambda_cPe^{-\frac{\lambda_a\sigma_a^2}{P}(2^{R_{S_2}^{II}}-1)}}{\lambda_cP\!+\!\lambda_a (Q_j\!-\!\Delta) (2^{R_{S_2}^{II}}\!-\!1)}\!\bigg)\\
\times\prod\limits_{i=1,i\neq j}^n\bigg(1-\frac{\lambda_cPe^{-\frac{\lambda_a\sigma_a^2}{P}(2^{R_{S_2}^{II}}-1)}}{\lambda_cP+\lambda_a Q_i (2_{S_2}^{R^{II}}-1)}\bigg)=\delta.\nonumber
\end{aligned}
\end{equation}

Similar to the proof of Proposition 3.2, $R_{S_2}^{II}$ is larger than $R_{S_1}^{II}$, while maintaining the same target outage probability $\delta$ at the receiver, but the non-outage probability at the monitor will decrease due to the higher transmission rate $R_{S_2}^{II}$.

From above, if the jamming power on any jammed channel decreases, the product of non-outage probability and non-own-goal probability at the monitor, i.e. the eavesdropping success probability, will degenerate. So the monitor will always use up all the jamming power.

Further since all channels are i.i.d. fading, by symmetry it is optimal to allocate the same amount of jamming power over $n$ jammed channels. Thus, if $n$ of $N$ channels are jammed, the monitor should evenly allocate all the jamming power over $n$ jammed channels, i.e.,  $Q_{\max}/n$.

\subsection{Proofs of Propositions 3.1 and 4.2}

We first prove that the own goal probability $\rho(n)$ increases as the number of jammed channels $n$ increases.

Assuming $k$ out of $N$ channels are jammed, the own goal probability can be expressed as
\begin{equation}
\begin{aligned}
\rho(k)=\mathbb{P}\bigg(\max\Big\{\frac{g_{a1}P}{g_{c1}\frac{Q_{\max}}{k}+\sigma_a^2},\cdots,\frac{g_{ak}P}{g_{ck}\frac{Q_{\max}}{k}+\sigma_a^2}\Big\}> \\ \max\Big\{\frac{g_{a1}P}{\sigma_a^2},\cdots, \frac{g_{a(N-k)}P}{\sigma_a^2}\Big\}\bigg).\nonumber
\end{aligned}
\end{equation}

If now $k+1$ channels are jammed, then the own goal probability becomes
\begin{equation}
\begin{aligned}
\rho(k\!+\!1)\!=\!\mathbb{P}\bigg(\!\!\max\Big\{\!\frac{g_{a1}P}{g_{c1}\frac{Q_{\max}}{k+1}\!+\!\sigma_a^2},\cdots,\frac{g_{a(k+1)}P}{g_{c(k+1)}\frac{Q_{\max}}{k+1}\!+\!\sigma_a^2}\!\Big\}> \\ \max\Big\{\frac{g_{a1}P}{\sigma_a^2},\cdots, \frac{g_{a(N-k-1)}P}{\sigma_a^2}\Big\}\!\bigg).\nonumber
\end{aligned}
\end{equation}

Since the number of jammed channels changes from $k$ to $k+1$, the jamming power on each jammed channel gets smaller, which makes them easier to be chosen by the ST for transmission. Also, there are more jammed channels (from $k$ to $k+1$) and less unjammed channels (from $N-k$ to $N-k-1$). Combining these two effects, clearly as the number of jammed channels $n$ increases, the own goal probability $\rho(n)$ will also increase.

Note that Proposition 3.1 is a special case of Proposition 4.2 and it is sufficient to prove Proposition 4.2 here. Denote $X_i=\frac{g_{ai}P}{\sigma_a^2}$ and $Y_i=\frac{g_{ai}P}{g_{ci}Q_i+\sigma_a^2}$, the CDFs of $X_i$ and $Y_i$ are given in \eqref{ii1} and \eqref{ii2}, respectively. Then we have
\begin{equation}
\begin{aligned}
\rho(n)\!=\!\!\int_{0}^{\infty}\!\!\mathbb{P}\{\max\{Y_{(1)},&\cdots,Y_{(n)}\geq x\}\} \\ &\times f_{\max\{X_{(1)},\cdots,X_{(N-n)}\}}(x)\,dx, \nonumber
\end{aligned}
\end{equation}where $f_{\max\{X_{(1)},\cdots,X_{(N-n)}\}}(x)$ is the PDF of the maximum SNR of the $N-n$ unjammed channels. Thus we have
\begin{equation}
\begin{aligned}
\rho(n) = \int_{0}^{\infty}\bigg(\int_{x}^{\infty} d[F_{Y}(y)]^n\bigg)d[F_{X}(x)]^{N-n} \nonumber
\end{aligned}
\end{equation}
\begin{equation}
\begin{aligned}
= \frac{(N-n)\lambda_a\sigma_a^2}{P}\int_{0}^{\infty} \bigg(1-(1-\frac{e^{-\frac{\lambda_a\sigma_a^2}{P}x}}{1+\frac{\lambda_aQ_{\max}}{n\lambda_cP}x})^n\bigg) \\ \times e^{-\frac{\lambda_a\sigma_a^2}{P}x}(1-e^{-\frac{\lambda_a\sigma_a^2}{P}x})^{N-n-1}dx \nonumber
\end{aligned}
\end{equation}
\begin{equation}
\begin{aligned}
\overset{\text{(a)}}{=} \frac{(N-n)\lambda_a\sigma_a^2}{P}\sum_{i=0}^{N-n-1}\sum_{j=1}^{n}{{N-n-1} \choose i}{{n} \choose j}(-1)^{i+j+1}\\ \times\int_{0}^{\infty} \frac{1}{(1+\frac{\lambda_aQ_{\max}}{n\lambda_cP}x)^j}e^{-(i+j+1)\frac{\lambda_a\sigma_a^2}{P}x}dx \nonumber
\end{aligned}
\end{equation}
\begin{equation}
\begin{aligned}
\overset{\text{(b)}}{=} \frac{(N-n)\lambda_a\sigma_a^2}{P}\sum_{i=0}^{N-n-1}\sum_{j=1}^{n}{{N-n-1} \choose i}{n \choose j}(-1)^{i+j+1}\\ \times\bigg(\frac{\lambda_aQ_{\max}}{n\lambda_cP}\bigg)^{-j} \bigg((1+i+j)\frac{\lambda_a\sigma_a^2}{P}\bigg)^{j-1}e^{\frac{(1+i+j)n\lambda_c\sigma_a^2}{Q_{\max}}}
\\ \times \Gamma\bigg(1-j,\frac{(1+i+j)n\lambda_c\sigma_a^2}{Q_{\max}}\bigg), \nonumber
\end{aligned}
\end{equation}where equality $\overset{\text{(a)}}{=}$ comes from the fact that the total number of channels $N$ and the number of jammed channels $n$ are both integers, following the binomial expansion of the two polynomial terms; and equality $\overset{\text{(b)}}{=}$ comes from reference [25, Eq. $3.353.2$], and $\Gamma(\cdot,\cdot)$ is the incomplete Gamma function. Note that $\Gamma(0,x)=-\text{Ei}(-x)$, which completes the proof of Proposition 3.1 with $N=2$ and $n=1$.

\subsection{Proof of Proposition 4.3}

Denote $u = 2^{R^{II}}-1$. We define the LHS of \eqref{iv17} as $h(n,u)$, which is the outage probability at the SR. Then we have
\begin{equation}
\begin{aligned}
h(n,u)-\delta=0. \nonumber
\end{aligned}
\end{equation}

By taking the first-order derivative of this implicit function over $u$ and $n$, we have the relationship between $u$ and $n$ as follows
\begin{equation}
\begin{aligned}
\frac{du}{dn}=-\frac{\partial h(n,u)/\partial n}{\partial h(n,u)/\partial u}. \nonumber
\end{aligned}
\end{equation}

We then can explicitly derive
\begin{equation}
\begin{aligned}
\frac{\partial h(n,u)}{\partial n}=\bigg(\frac{-\frac{\lambda_aQ_{\max}}{\lambda_cP}\frac{u}{n}e^{-\frac{\lambda_a\sigma_a^2}{P}u}}{(1+\frac{\lambda_aQ_{\max}}{\lambda_cP}\frac{u}{n})^2(1-\frac{e^{-\frac{\lambda_a\sigma_a^2}{P}u}}{1+\frac{\lambda_aQ_{\max}}{\lambda_cP}\frac{u}{n}})}\\ +\ln\Big(1+\frac{\frac{\frac{\lambda_aQ_{\max}}{\lambda_cP}\frac{u}{n}}{1+\frac{\lambda_aQ_{\max}}{\lambda_cP}\frac{u}{n}}e^{-\frac{\lambda_a\sigma_a^2}{P}u}}{1-e^{-\frac{\lambda_a\sigma_a^2}{P}u}}\Big)\bigg),\nonumber
\end{aligned}
\end{equation}and according to inequality $\ln(1+x)>\frac{x}{1+x}$ for any $x>0$, we have
\begin{equation}
\begin{aligned}
\frac{\partial h(n,u)}{\partial n}\!>\!\frac{\frac{\lambda_aQ_{\max}}{\lambda_cP}\frac{u}{n}e^{-\frac{\lambda_a\sigma_a^2}{P}u}}{(1\!+\!\frac{\lambda_aQ_{\max}}{\lambda_cP}\frac{u}{n}\!)(1\!-\!\frac{e^{-\frac{\lambda_a\sigma_a^2}{P}u}}{1+\frac{\lambda_aQ_{\max}}{\lambda_cP}\frac{u}{n}}\!)} \!\bigg(\!1\!-\!\frac{1}{1\!\!+\!\!\frac{\lambda_aQ_{\max}}{\lambda_cP}\frac{u}{n}}\!\bigg)\\>0.
\label{ap22}
\end{aligned}
\end{equation}

For $\frac{\partial h(n,u)}{\partial u}$, as the transmission rate $R^{II}$ increases, outage probability at the SR $h(n,u)$ increases, thus
\begin{equation}
\begin{aligned}
\frac{\partial h(n,u)}{\partial u}>0.
\label{ap23}
\end{aligned}
\end{equation}

Combining \eqref{ap22} and \eqref{ap23}, we have
\begin{equation}
\begin{aligned}
\frac{du}{dn}=-\frac{\partial h(n,u)/\partial n}{\partial h(n,u)/\partial u}<0. \nonumber
\end{aligned}
\end{equation}

By substituting $R^{II}=\log_2(1+u)$ back, we have
\begin{equation}
\begin{aligned}
\frac{dR^{II}}{dn}=\frac{du}{dn}\frac{dR^{II}}{du}<0. \nonumber
\end{aligned}
\end{equation}

Thus, as the number of jammed channels $n$ increases, the transmission rate $R^{II}(n)$ decreases.

\subsection{Proof of Proposition 5.1}

Consider the expression of eavesdropping success probability $\varphi^{II}(n)$ in \eqref{iv15}, which is the product of non-own goal probability $1-\rho(n)$ multiplied by non-outage probability at the monitor $e^{-\frac{\lambda_b\sigma_b^2}{P}(2^{R^{II}(n)}-1)}$.

For any fixed $n$, as jamming power budget $Q_{\max}$ increases, the first part, non-own goal probability, $1-\rho(n)$ increases, since the jammed channels become less likely to be chosen by the suspicious link. The second part, non-outage probability at the monitor, $e^{-\frac{\lambda_b\sigma_b^2}{P}(2^{R^{II}(n)}-1)}$ also increases, since the suspicious link will transmit at a lower rate to maintain target outage probability and the monitor can eavesdrop more clearly. Thus $\varphi^{II}(n)$ monotonically increases with $Q_{\max}$ for any fixed $n$.

For $\varphi_{AB}^{II}(n)$, we can see from \eqref{iv20} that
\begin{equation}
\begin{aligned}
\lim_{Q_{\max} \to 0^+}\varphi_{AB}^{II}(1)>\lim_{Q_{\max} \to 0^+}\varphi_{AB}^{II}(2), \nonumber
\end{aligned}
\end{equation}and from \eqref{iv21}, we can see that
\begin{equation}
\begin{aligned}
\lim_{Q_{\max} \to \infty}\varphi_{AB}^{II}(1)<\lim_{Q_{\max} \to \infty}\varphi_{AB}^{II}(2). \nonumber
\end{aligned}
\end{equation}

Thus there exists a unique solution $Q_{\max}$, so that
\begin{equation}
\begin{aligned}
\varphi_{AB}^{II}(1)=\varphi_{AB}^{II}(2), \nonumber
\end{aligned}
\end{equation}and we call this solution as $\underline{Q}_{AB}$.

By assuming $\varphi_{AB}^{II}(n)$ is unimodal or monotonic in $n$, if $Q_{\max}<\underline{Q}_{AB}$, from $\varphi_{AB}^{II}(1)>\varphi_{AB}^{II}(2)$ we can conclude that $\varphi_{AB}^{II}(n)$ now monotonically decreases with $n$. For $\varphi_{BA}^{II}(n)$, we can derive $\underline{Q}_{BA}$ so that when $Q_{\max}<\underline{Q}_{BA}$, $\varphi_{BA}^{II}(n)$ decreases with $n$. Define $\underline{Q}=\min\{\underline{Q}_{AB},\underline{Q}_{BA}\}$, when $Q_{\max}<\underline{Q}$, both $\varphi_{AB}^{II}(n)$ and $\varphi_{BA}^{II}(n)$ decrease with $n$, thus it is optimal for the monitor to jam $n^*=1$ channel.

Similarly for $\varphi_{AB}^{II}(n)$, we can see from \eqref{iv20} that
\begin{equation}
\begin{aligned}
\lim_{Q_{\max} \to 0^+}\varphi_{AB}^{II}(N-2)>\lim_{Q_{\max} \to 0^+}\varphi_{AB}^{II}(N-1), \nonumber
\end{aligned}
\end{equation}and from \eqref{iv21}, we can see that
\begin{equation}
\begin{aligned}
\lim_{Q_{\max} \to \infty}\varphi_{AB}^{II}(N-2)<\lim_{Q_{\max} \to \infty}\varphi_{AB}^{II}(N-1). \nonumber
\end{aligned}
\end{equation}

There exists a unique solution $Q_{\max}$, so that
\begin{equation}
\begin{aligned}
\varphi_{AB}^{II}(N-2)=\varphi_{AB}^{II}(N-1), \nonumber
\end{aligned}
\end{equation}and we call this solution as $\bar{Q}_{AB}$.

By assuming $\varphi_{AB}^{II}(n)$ is unimodal or monotonic in $n$, if $Q_{\max}>\bar{Q}_{AB}$, from $\varphi_{AB}^{II}(N-2)<\varphi_{AB}^{II}(N-1)$ we can conclude that $\varphi_{AB}^{II}(n)$ now monotonically increases with $n$. Similarly for $\varphi_{BA}^{II}(n)$, we can derive $\bar{Q}_{BA}$ so that when $Q_{\max}>\bar{Q}_{BA}$, $\varphi_{BA}^{II}(n)$ increases with $n$. Define $\bar{Q}=\max\{\bar{Q}_{AB},\bar{Q}_{BA}\}$, when $Q_{\max}<\bar{Q}$, both $\varphi_{AB}^{II}(n)$ and $\varphi_{BA}^{II}(n)$ increase with $n$, thus it is optimal for the monitor to jam $n^*=N-1$ channels.

When $\underline{Q}\leq Q_{\max}\leq \bar{Q}$, at least one of $\varphi_{AB}^{II}(n)$ and $\varphi_{BA}^{II}(n)$ is unimodal. Without loss of generality, we assume $n_{AB}^*\leq n_{BA}^*$, clearly when $n< n_{AB}^*$, both $\varphi_{AB}^{II}(n)$ and $\varphi_{BA}^{II}(n)$ are monotonically increasing, and when $n> n_{BA}^*$, both $\varphi_{AB}^{II}(n)$ and $\varphi_{BA}^{II}(n)$ are monotonically decreasing. Thus $n^*$ must lie between $n_{AB}^*$ and $n_{BA}^*$, and can be numerically searched.

%

\begin{IEEEbiography}[{\includegraphics[width=1in,height=1.25in,clip,keepaspectratio]{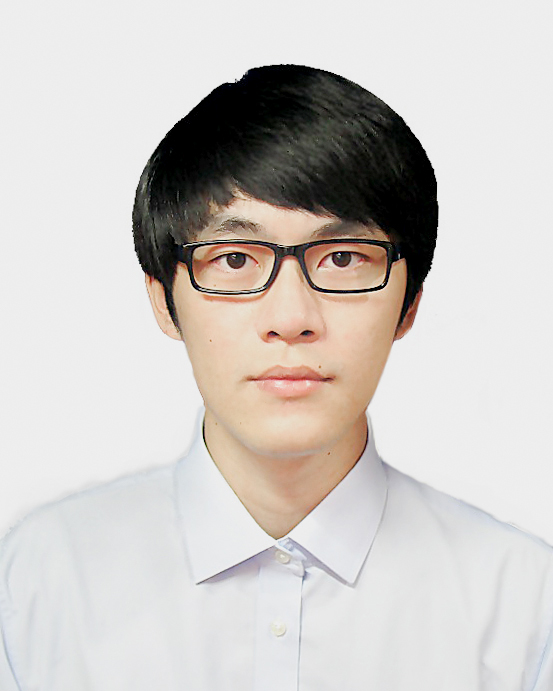}}]{Yitao Han}
(S'19) received the B. Eng. degree in information and communication engineering from Zhejiang University, Hangzhou, China, in 2015 and the M. Sc. degree in Department of Electrical and Computer Engineering from National University of Singapore, Singapore, in 2016. He is currently pursuing the Ph. D. degree in Engineering System and Design pillar, Singapore University of Technology and Design, Singapore, under SUTD-NUS joint Ph. D. programme with President's Graduate Fellowship.

His current research interests include physical layer security and UAV communications.
\end{IEEEbiography}

\begin{IEEEbiography}[{\includegraphics[width=1in,height=1.25in,clip,keepaspectratio]{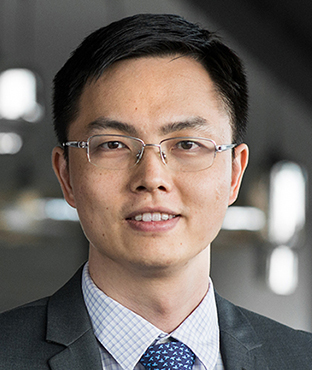}}]{Lingjie Duan}
(S'09-M'12-SM'16) received the Ph.D. degree from The Chinese University of Hong Kong in 2012. He is currently an Assistant Professor with the Engineering Systems and Design Pillar, Singapore University of Technology and Design, Singapore. His current research interests include network economics and game theory, cognitive and cooperative communications, energy harvesting wireless communications, mobile crowdsourcing, and wireless information surveillance. He was a recipient of the 2016 SUTD Excellence in Research Award, and the 10th IEEE ComSoc Asia-Pacific Outstanding Young Researcher Award in 2015. He was also the Finalist of the Hong Kong Young Scientist Award 2014 under Engineering Science track. He is now an Editor of the IEEE TRANSACTIONS ON WIRELESS COMMUNICATIONS and IEEE COMMUNICATIONS SURVEYS AND TUTORIALS. In 2016, he was a Guest Editor of the IEEE JOURNAL ON SELECTED AREAS IN COMMUNICATIONS Special Issue on Human-in-the-loop Mobile Networks, and was also a Guest Editor of the IEEE Wireless Communications Magazine for feature topic Sustainable Green Networking and Computing in 5G Systems. He is also a regular TPC member of a number of leading conferences on wireless communications and networking (e.g., IEEE INFOCOM, SECON, WiOPT, GLOBECOM, ICC and ACM MobiHoc).
\end{IEEEbiography}

\begin{IEEEbiography}[{\includegraphics[width=1in,height=1.25in,clip,keepaspectratio]{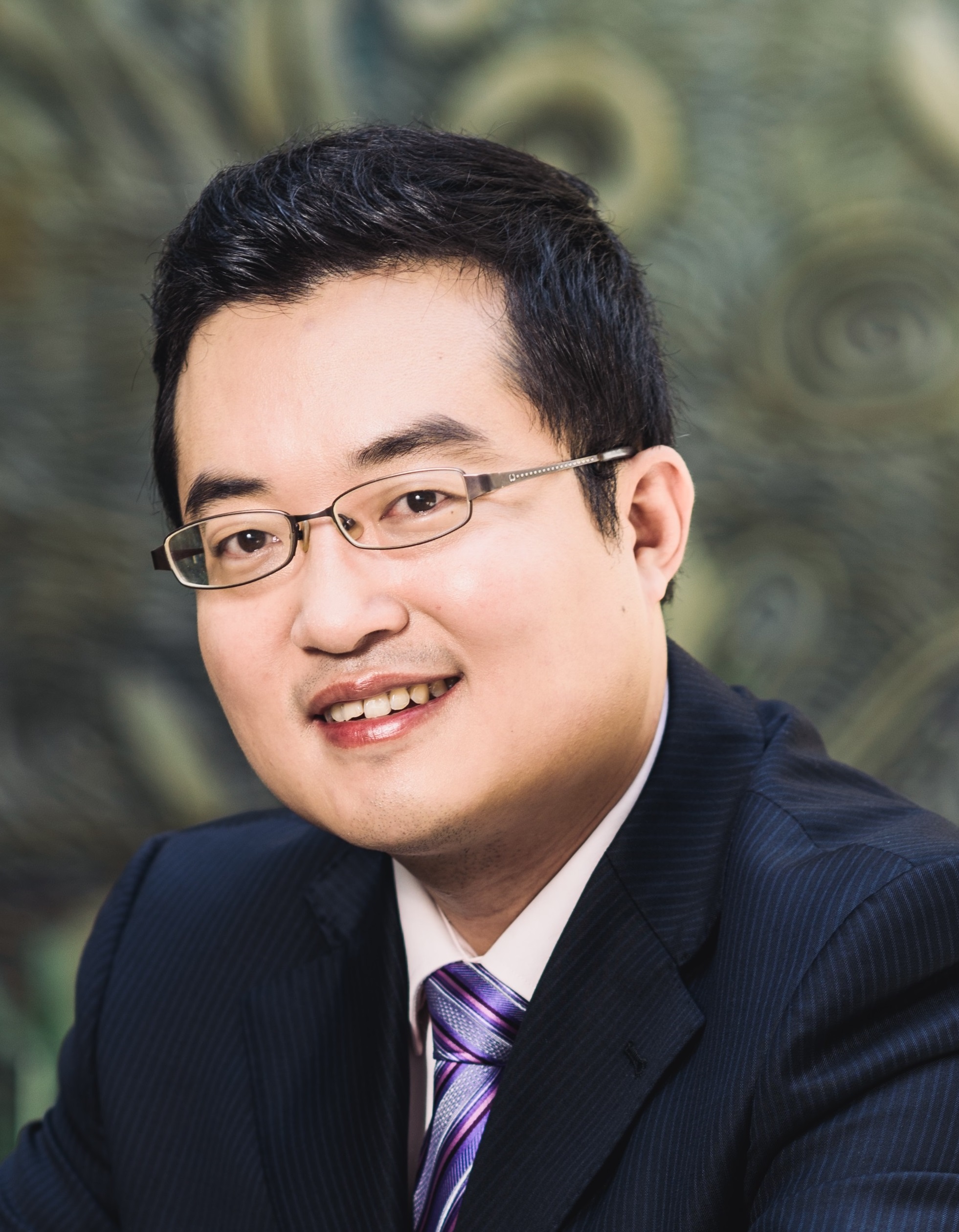}}]{Rui Zhang}
(S'00-M'07-SM'15-F'17) received the B.Eng. (first-class Hons.) and M.Eng. degrees from the National University of Singapore, Singapore, and the Ph.D. degree from the Stanford University, Stanford, CA, USA, all in electrical engineering.

From 2007 to 2010, he worked as a Research Scientist with the Institute for Infocomm Research, ASTAR, Singapore. Since 2010, he has joined the Department of Electrical and Computer Engineering, National University of Singapore, where he is now a Dean's Chair Associate Professor in the Faculty of Engineering. He has authored over 300 papers. He has been listed as a Highly Cited Researcher (also known as the World's Most Influential Scientific Minds), by Thomson Reuters (Clarivate Analytics) since 2015. His research interests include UAV/satellite communication, wireless information and power transfer, multiuser MIMO, smart and reconfigurable environment, and optimization methods.

He was the recipient of the 6th IEEE Communications Society Asia-Pacific Region Best Young Researcher Award in 2011, and the Young Researcher Award of National University of Singapore in 2015. He was the co-recipient of the IEEE Marconi Prize Paper Award in Wireless Communications in 2015, the IEEE Communications Society Asia-Pacific Region Best Paper Award in 2016, the IEEE Signal Processing Society Best Paper Award in 2016, the IEEE Communications Society Heinrich Hertz Prize Paper Award in 2017, the IEEE Signal Processing Society Donald G. Fink Overview Paper Award in 2017, and the IEEE Technical Committee on Green Communications \& Computing (TCGCC) Best Journal Paper Award in 2017. His co-authored paper received the IEEE Signal Processing Society Young Author Best Paper Award in 2017. He served for over 30 international conferences as the TPC co-chair or an organizing committee member, and as the guest editor for 3 special issues in the IEEE JOURNAL OF SELECTED TOPICS IN SIGNAL PROCESSING and the IEEE JOURNAL ON SELECTED AREAS IN COMMUNICATIONS. He was an elected member of the IEEE Signal Processing Society SPCOM Technical Committee from 2012 to 2017 and SAM Technical Committee from 2013 to 2015, and served as the Vice Chair of the IEEE Communications Society Asia-Pacific Board Technical Affairs Committee from 2014 to 2015. He served as an Editor for the IEEE TRANSACTIONS ON WIRELESS COMMUNICATIONS from 2012 to 2016, the IEEE JOURNAL ON SELECTED AREAS IN COMMUNICATIONS: Green Communications and Networking Series from 2015 to 2016, and the IEEE TRANSACTIONS ON SIGNAL PROCESSING from 2013 to 2017. He is now an Editor for the IEEE TRANSACTIONS ON COMMUNICATIONS and the IEEE TRANSACTIONS ON GREEN COMMUNICATIONS AND NETWORKING. He serves as a member of the Steering Committee of the IEEE Wireless Communications Letters. He is a Distinguished Lecturer of IEEE Signal Processing Society and IEEE Communications Society.
\end{IEEEbiography}

\end{document}